\pdfoutput=1%
\documentclass[%
times,%
final,%
5p,%
twocolumn,%
sort&compress,%
10pt%
]{elsarticle}

\journal{Physica A: Statistical Mechanics and its Applications}

\biboptions{square,comma,sort&compress}

\usepackage{xcolor,colortbl}
\definecolor{oxfordblue}{rgb}{0.0, 0.13, 0.28}
\definecolor{forestgreenTrad}{rgb}{0.0, 0.27, 0.13}
\definecolor{harvardcrimson}{rgb}{0.65, 0.0, 0.09}
\usepackage{graphicx}
\usepackage{dcolumn}
\usepackage{bm}
\usepackage[normalem,normalbf]{ulem}
\usepackage{enumerate}
\usepackage{tikz,pgfplots}
\pgfplotsset{compat=newest}
\usepackage[english]{babel}
\usepackage[utf8]{inputenc}
\usepackage[T1]{fontenc}
\usepackage{amsmath}
\usepackage[%
	activate={true,nocompatibility},%
	final,%
	tracking=true,%
	kerning=true,%
	spacing=true,%
	factor=1100,%
	stretch=10,%
	shrink=10]{microtype}
\microtypecontext{spacing=nonfrench}
\usepackage{enumitem}
\usepackage[colorlinks]{hyperref}
\AtBeginDocument{%
	\hypersetup{%
    	colorlinks = true,%
    	linkcolor  = oxfordblue,%
    	citecolor  = harvardcrimson,%
    	urlcolor   = forestgreenTrad%
    }%
}%
\usepackage[capitalize]{cleveref}
\usepackage{url}
\PassOptionsToPackage{hyphens}{url}
\usepackage[caption=true]{subfig}
\AtBeginDocument{%
	\captionsetup[figure]{%
		font=scriptsize,%
		labelfont={bf,it},%
		textfont=it,%
		justification=justified,%
		singlelinecheck=true,%
		format=plain,%
		name=Fig. ,%
		labelsep=period}
	\captionsetup[table]{%
		font=scriptsize,%
		labelfont={bf,it},%
		textfont=it,%
		justification=justified,%
		singlelinecheck=true,%
		format=plain,%
		name=Table ,%
		labelsep=period}
}%
\usepackage{afterpage}
\usepackage{placeins}
\usepackage{booktabs}
\usepackage{multirow}
\usepackage{csquotes}
\usepackage{grffile}
\usepackage{array}
\newcolumntype{L}[1]{>{\raggedright\let\newline\\\arraybackslash\hspace{0pt}}m{#1}}
\newcolumntype{C}[1]{>{\centering\let\newline\\\arraybackslash\hspace{0pt}}m{#1}}
\newcolumntype{R}[1]{>{\raggedleft\let\newline\\\arraybackslash\hspace{0pt}}m{#1}}
\newcommand{\bc}{\begin{center}}
\newcommand{\ec}{\end{center}}
\newcommand{\be}{\begin{equation}}
\newcommand{\ee}{\end{equation}}
\newcommand{\bea}{\begin{eqnarray}}
\newcommand{\eea}{\end{eqnarray}}
\newcommand{\beq}{\begin{eqnarray*}}
\newcommand{\eeq}{\end{eqnarray*}}

\begin{document}

\fontsize{9pt}{11pt}\selectfont%


\begin{frontmatter}

\title{Deviations in expected price impact\\for  small transaction volumes under fee restructuring}

\author[wits]{M.~Harvey\corref{cor1}}
\ead{michael.harvey@students.wits.ac.za}
\author[wits]{D.~Hendricks}
\author[wits]{T.~Gebbie}
\author[wits]{D.~Wilcox}

\journal{Physica A}

\address[wits]{School of Computer Science and Applied Mathematics, University of the Witwatersrand, Johannesburg, South Africa}

\cortext[cor1]{Corresponding author}

\begin{abstract}
We report on the occurrence of an anomaly in the price impacts of small transaction volumes following a change in the fee structure of an electronic market. We first review evidence for the existence of a master curve for price impact on the Johannesburg Stock Exchange (JSE). On attempting to re-estimate a master curve after fee reductions, it is found that the price impact corresponding to smaller volume trades is greater than expected relative to prior estimates for a range of listed stocks. We show that a master curve for price impact can be found following rescaling by an appropriate liquidity proxy, providing a means for practitioners to approximate price impact curves without onerous processing of tick data.
\end{abstract}


\end{frontmatter}

\section{Introduction}\label{intro}

Large market orders submitted to electronic trading platforms have the tendency to deplete more than the inventory available at the best bid/offer level at time of transaction. This provides a mechanistic account for change in prevailing asset prices. Observation of this phenomenon  has translated to various views on the definition and estimation of the impact of a single transaction as a component of transaction cost, where the latter includes easily measurable fees and commissions as well as indirect levies such as the component of price impact.

With trade volumes increasing 10-fold in developed markets through the 1990's, it became possible to quantify regularities in price response to establish the relationship between averaged price shifts and transaction volumes~%
\cite{%
Lillo2003,%
Iori2003,%
Potters2003,%
Plerou2004,%
Farmer2004,%
Zawadowski2004,%
BGPW2004,%
Moro2009,%
Gatheral2009,%
Wilinski2015}. This has led to a more refined but heuristic definition of price impact as {\em the correlation between trade size and direction and the resultant price change}.

In a study of 1000 stocks listed on the New York Stock Exchange for 1995-1998, it was determined that inclusion of an adjustment for the liquidity variation for stocks in different  market capitalisation classes provided a calibration of a single representative price impact function for the entire market~\cite{Lillo2003}.

Measuring  by market capitalisation, the Johannesburg Stock Exchange (JSE) is one of the top 20 stock exchanges in the world, where current JSE market capitalisation of 1 trillion USD is approximately $1/20$\textsuperscript{th} of the NYSE. After being located in London since 2001, the JSE trade infrastructure was repatriated in July 2012~\cite{JSE:MilleniumITGoLiveNotice2012}. In September 2013, an overhaul of the JSE equity trading fee structure\footnote{The term `fee structure' refers to the `JSE Equity Market Transaction Billing Model' as discussed in~\cite{JSE:EquityMarketBillingModelChangeNotice2013}.}~resulted in the dramatic reduction of transaction costs~\cite{%
JSE:EquityMarketPriceList2013,%
JSE:EquityMarketBillingModelChangeNotice2013,%
JSE:EquityMarketPriceList2014v1a}.~%
In particular, a minimum fee of 4.00 ZAR per trade (excl. VAT) was eliminated so that all transaction fees were reduced to pro rata amounts, depending linearly on transaction value (up until a capped maximum charge of 300 ZAR (excl. VAT))~\cite{%
JSE:EquityMarketPriceList2013,%
JSE:EquityMarketBillingModelChangeNotice2013,%
JSE:EquityMarketPriceList2014v1a}.
\begin{table}[tbp]
\caption{%
Dates of market structure events on the JSE. EVENT I refers to the date when the JSE's current trading system, Millenium Exchange, went live. This change involved moving the server from London to Johannesburg. EVENT II refers to an equity trading fee structure change. EVENT III refers to the date on which JSE Colocation Services went live~\cite{%
JSE:MilleniumITGoLiveNotice2012,%
JSE:EquityMarketPriceList2013,%
JSE:EquityMarketBillingModelChangeNotice2013,%
JSE:EquityMarketPriceList2014v1a,%
JSE:ColocationGoesLive2014,%
JSE:ColocationBrochure2014}.}\label{JSEdatesTable}
\centering
\footnotesize
\begin{tabular}{@{}lll@{}}
\toprule
\textbf{Name} 	& \textbf{Date} & \textbf{Description}  \\ \midrule
EVENT I 		& 02-Jul-2012 	& JSE MillenniumIT server relocation \\
EVENT II 		& 30-Sep-2013 	& JSE equity trading fee structure change \\
EVENT III 		& 12-May-2014 	& JSE Colocation Services go live\\ \bottomrule
\end{tabular}
\end{table}
\Cref{JSEdatesTable}~provides a time line of three key exchange infrastructure changes since Jan-2012~\cite{%
JSE:MilleniumITGoLiveNotice2012,%
JSE:EquityMarketPriceList2013,%
JSE:EquityMarketBillingModelChangeNotice2013,%
JSE:EquityMarketPriceList2014v1a,%
JSE:ColocationGoesLive2014,%
JSE:ColocationBrochure2014}.~%
As anticipated,  the fee structure change has corresponded to measurable changes to the JSE market microstructure. In this investigation we report the effect of the fee change on the price impact curves for 2013.  Market microstructure and price response for stocks listed on the JSE in 2011 were investigated in~\cite{Dupreez}~and optimal order execution on the JSE to minimise indirect costs was investigated in~\cite{Hendricks2014}.

The asymmetrical impact of buyer and seller initiated transactions and the effect of transaction duration on impact is investigated in~\cite{Yang2011}.

Our analysis highlights that the regulatory effect of  direct transaction costs are coupled to nonlinear sensitivities in price impact of trade. In a different study \cite{Malinova2011}, it was found that the introduction of liquidity rebates on the Toronto Stock Exchange did not cause a decrease in trading costs for market orders but did result in a revenue increase for liquidity providers. The capacity to measure such feedbacks supports risk measurement and the reduction of some investor uncertainty from a market microstructure perspective.

The paper is structured as follows: in~\cref{priceimpact}~we give a brief review of the price impact master curve conjecture which is tested in this paper for the three major sectors of the JSE Top 40. Data and curve construction are reviewed in~\cref{data}, results are documented in~\cref{results} and we conclude with remarks on implications.

\section{Price impact}\label{priceimpact}

Several studies have scrutinised components of price impact, where the latter includes both permanent and transient components, and comprehensive reviews are included in~\cite{%
Farmer2004,%
BGPW2004,%
Moro2009,%
Gatheral2009,%
Wilinski2015}. A reasonable  point of departure is the identification of the following very general relationship~\cite{Potters2003}:
{\em price impact quantifies how a transaction of a given volume affects the price}.

Lillo {\em et al.}~\cite{Lillo2003}~focus on the relationship between transaction volume and the immediate price increment which follows. Letting $p(t)$ be the logarithm of the midquote price, the impact of a transaction occurring at time $t$ is defined as the increment:
\begin{align*}
\Delta p(t_{k+1}) = p(t_{k+1})-p(t_k)
\end{align*}
where $t_k$ and $t_{k+1}$ respectively indicate the times of the quotes that precede and immediately follow the transaction.

They conjecture a power-law relationship between transaction size and price change for subsets of stocks grouped by market capitalisation, with transaction direction accounted for by partitioning trades into buyer and seller initiated orders. For transactions inducing an average price shift of $\Delta p^{\ast}$ the following relationship was verified:~%
\begin{align}
\Delta p^{\ast} = \frac{\text{sign}(\omega^{\ast})\vert\omega^{\ast}\vert^{\alpha}}{\lambda},\label{eq:PLawConjecture}%
\end{align}
where $\omega^{\ast}$ denotes average normalised transaction size in number of shares and $\lambda$ denotes a liquidity parameter.

In their investigation~\cite{Lillo2003}~the quantity $C^{0.4}$ was found to be a good proxy for liquidity, where $C$ denotes the average market capitalisation\footnote{This investigation makes the pragmatic choice of using average daily value traded for each group over a given period as the proxy for liquidity. This is different to~\cite{Lillo2003}~where the average market capitalisation of each group over a given period is used as the proxy for liquidity.}. To obtain a general representation of the price impact function, the following price law was conjectured:%
\begin{align}
\Delta p^{\ast}(\omega^{\ast},C) = C^{-\gamma}f(\omega^{\ast} C^{\delta}).\label{eq:MasterPIFunction}%
\end{align}
To identify a linear dependence structure,~%
$\omega^{\ast}$%
~and~%
$\Delta p^{\ast}$%
~were rescaled by~%
$\omega^{\ast} \rightarrow {\omega^{\ast}}/{C^{\delta}}$~%
and~%
$\Delta p^{\ast} \rightarrow \Delta p^{\ast} C^{\gamma}$.~%
Calibration dependence in~\cite{Lillo2003}~yielded parameter values of $\delta \approx \gamma \approx 0.3$.

We tested this conjecture for three key sectors listed on the JSE for 2013.

\section{Data and curve construction}\label{data}

This investigation is based on Thompson Reuters Tick History trades and quotes data for the JSE Top 40 stocks for the period 01-Jan-2013 to 31-Dec-2013.

The constituents of the JSE Top 40 were grouped according to the SA sector classification into three sectors: Financials (JSE-FINI), Resources (JSE-RESI) and Industrials (JSE-INDI). The constituents of each sector are given in~\ref{app:sectorconstituents}. We chose to consider sector grouping, since the JSE market capitalisation is dominated by Resource companies \cite{CU2004}, which would bias market capitalisation based groupings. Price impact curves were computed for:

\begin{itemize}
   \item the period before the fee structure change: 01-Jan-2013 (Tuesday) to 27-Sep-2013 (Friday)~%
   \cite{JSE:EquityMarketPriceList2013},
   \item the period after the fee structure change: 30-Sep-2013 (Monday) to 31-Dec-2013 (Tuesday)~%
   \cite{JSE:EquityMarketBillingModelChangeNotice2013,JSE:EquityMarketPriceList2014v1a}.
\end{itemize}

A total of 250 trading days were analysed with the periods before and after the fee structure change accounting for 186 and 64 days respectively. The data was filtered to exclude price formation during auction periods and the more volatile 10-minute periods at commencement of regular trade (09h00-09h10) and before the closing  auction (16h50-17h00), as well as nonsensical records such as trades and quotes with zero volume or zero price\footnote{Transactions with questionably large traded volumes (in excess of an order of magnitude of 6) were assumed to be off-market trades and were therefore removed from the data set.}.

To facilitate the identification of the associated midquote price response to a trade of size $\omega$, all trade events that had the same date-time stamp were aggregated into single events by computing volume weighted average prices (VWAP) and cumulative volumes. For quote events that had the same date-time stamp the last quote was used. This resulted in a processed dataset with unique microsecond resolution time-steps, where the time between events may be irregular. The impact of trade direction was taken into account by partitioning the data into buyer and seller initiated trades. Since most limit order book data does not include information regarding trade direction, the Lee-Ready~\cite{LeeReady1991} algorithm was adopted to infer this.

For each trade record occurring at time $t_k$, the associated price shift $\Delta p(t_k)$ is computed as in~Lillo {\em et al.}~\cite{Lillo2003}~where $p(t_k)$ and $p(t_{k-1})$ are respectively the log-midquote price immediately prior and immediately after the trade event. This change in the log-midquote price quantifies immediate price response to a trade of volume $\omega(t_k)$.

To facilitate comparison between stocks the volumes were normalised by dividing through by the average volume of each stock\footnote{From this point on $\omega(t_k)$ is taken to be the normalised volume of a transaction occurring at time $t_k$.}. Transactions resulting in no price change (i.e. $\Delta p = 0$) were included in the analysis.

Next, trade records $\left( \omega(t_k), \Delta p(t_k) \right)$ for each time $t_k$ were binned according to normalised trade volume by constructing 20 logarithmically spaced bins between the normalised trade volumes of $10^{-3.2}$ and $10$. For each volume-bin, the average price impact $\Delta p^{\ast}$ and average normalised volume $\omega^{\ast}$ were computed. Results were plotted on a log-log scale.

\section{Results}\label{results}

\begin{figure*}[tbp!]
\centering
\subfloat[Period before fee structure change.]{%
\includegraphics[width=0.49\textwidth]{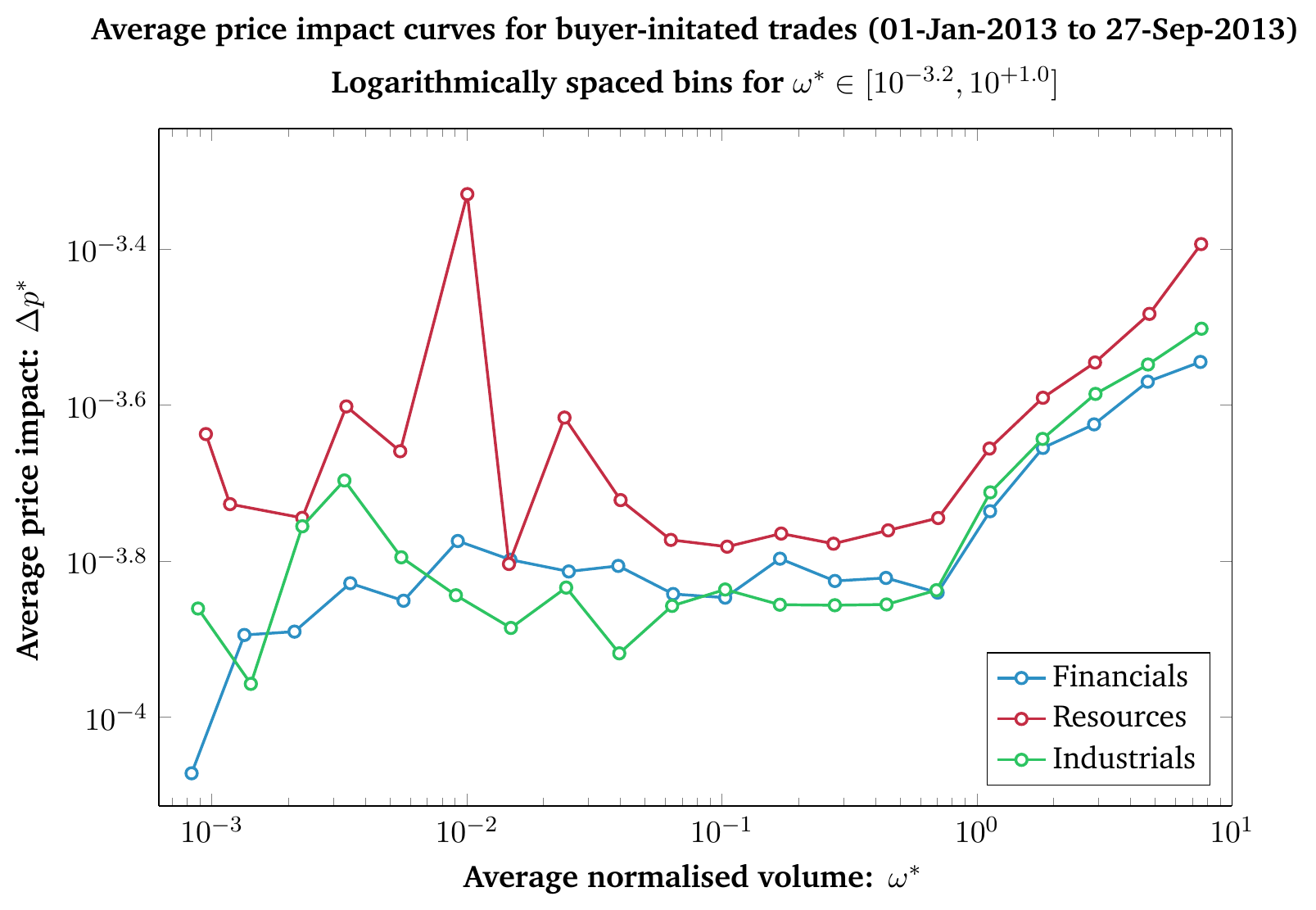}%
}
\hfill
\subfloat[Period after fee structure change.]{%
\includegraphics[width=0.49\textwidth]{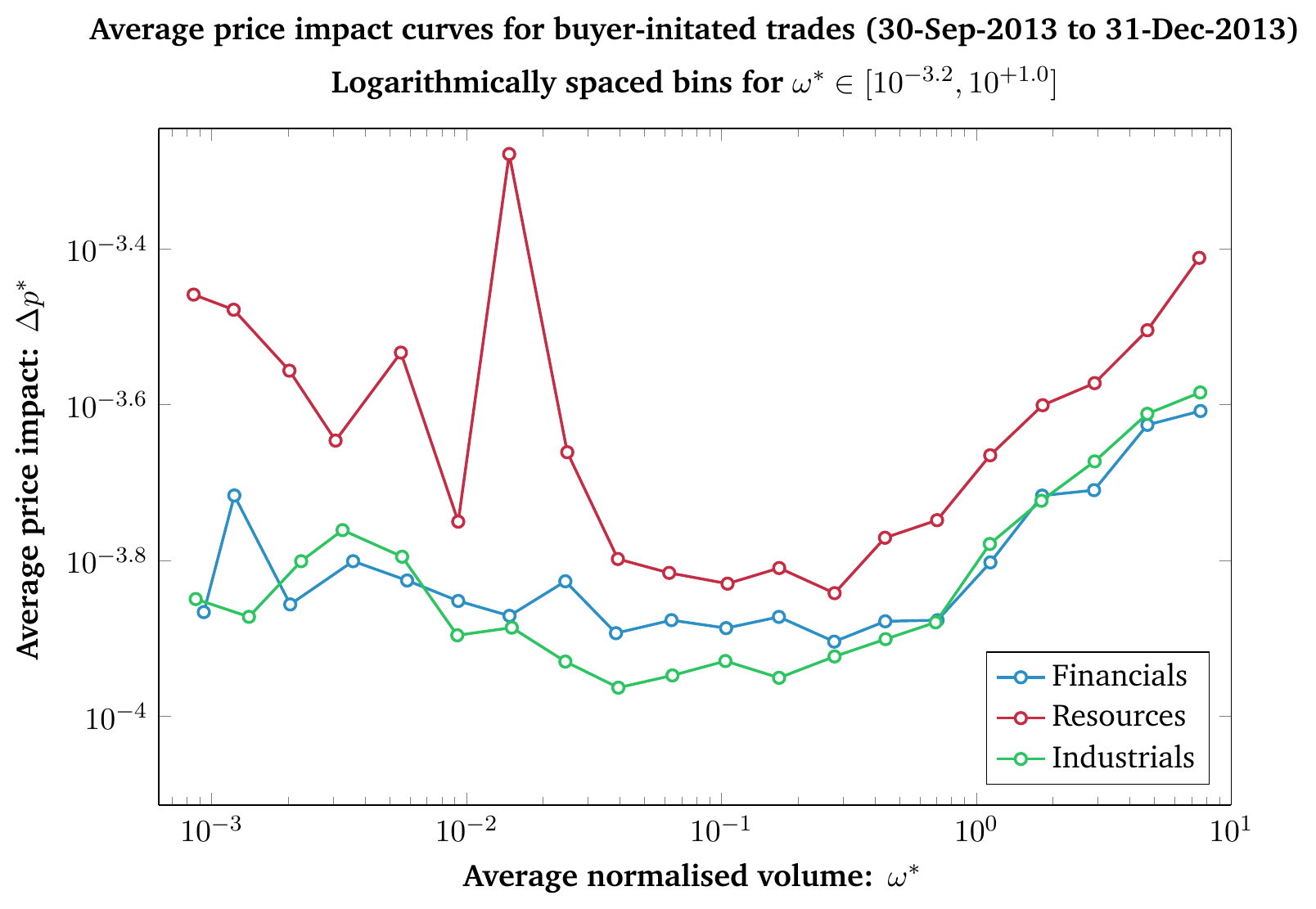}%
}
\caption{%
Average price impact curves for buyer initiated transactions of constituents of the Financials (JSE-FINI), Resources (JSE-RESI) and Industrials (JSE-INDI) sectors for the periods 01-Jan-2013 to  27-Sep-2013 (left) and 30-Sep-2013 to 31-Dec-2013 (right).
}\label{GroupPICurvesBuyer}
\end{figure*}

\begin{figure*}[tbp!]
\centering
\subfloat[Period before fee structure change.]{%
\includegraphics[width=0.49\textwidth]{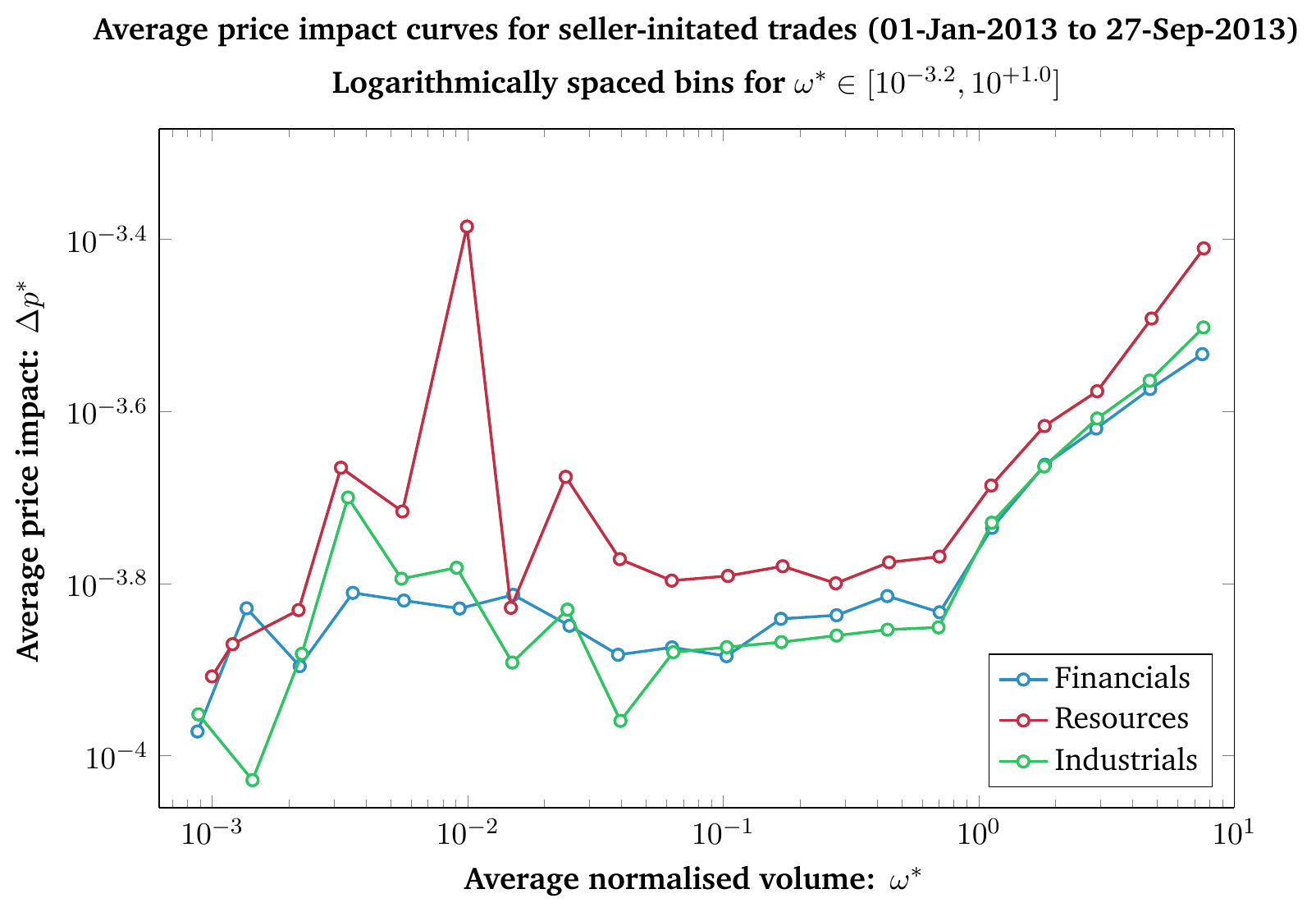}%
}
\hfill
\subfloat[Period after fee structure change.]{%
\includegraphics[width=0.49\textwidth]{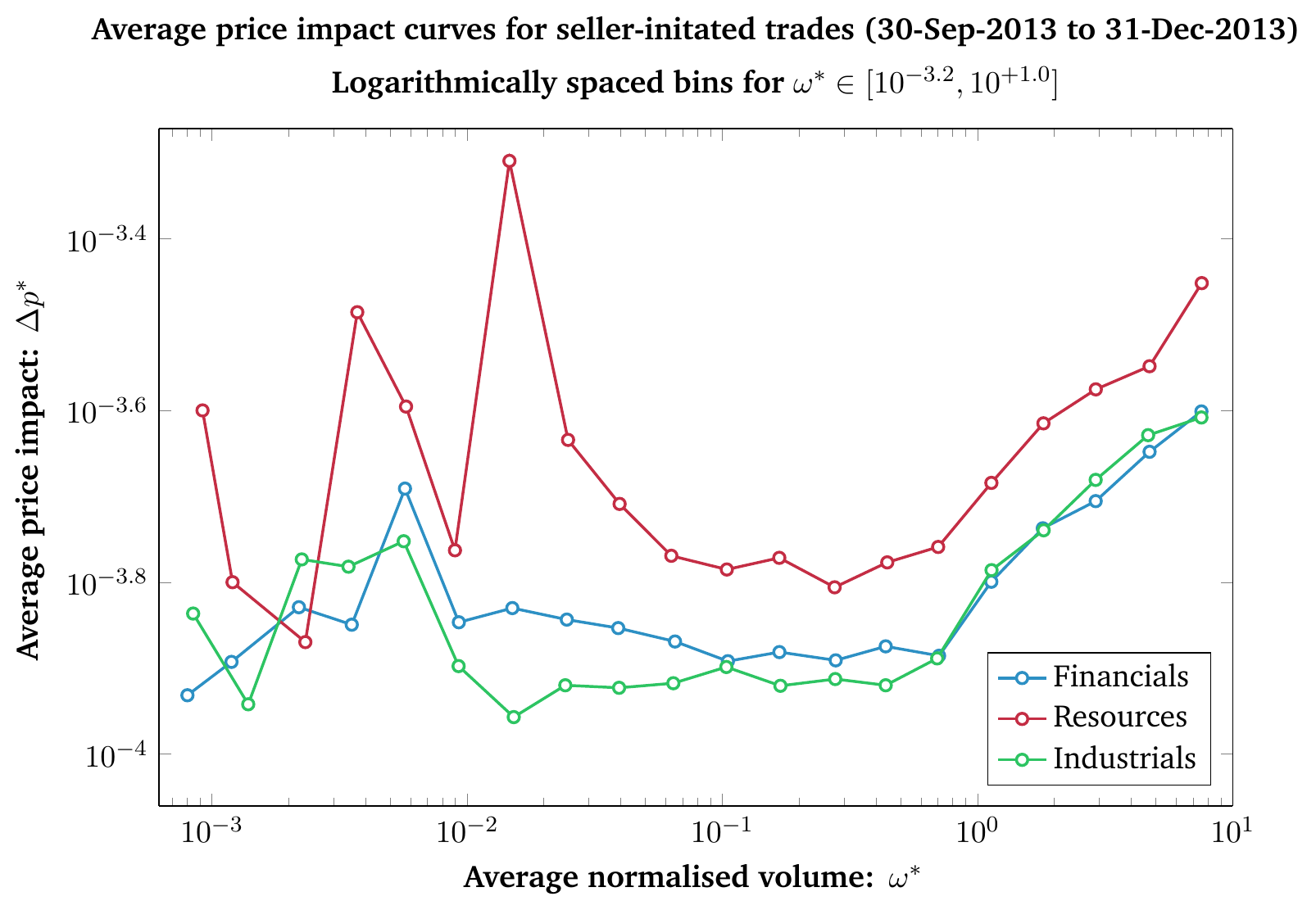}%
}
\caption{%
Average price impact curves for seller initiated transactions of constituents of the Financials (JSE-FINI), Resources (JSE-RESI) and Industrials (JSE-INDI) sectors for the periods 01-Jan-2013 to  27-Sep-2013 (left) and 30-Sep-2013 to 31-Dec-2013 (right).
}\label{GroupPICurvesSeller}
\end{figure*}

\begin{figure*}[tbp!]
\centering
\subfloat[Period before fee structure change.]{%
\includegraphics[width=0.49\textwidth]{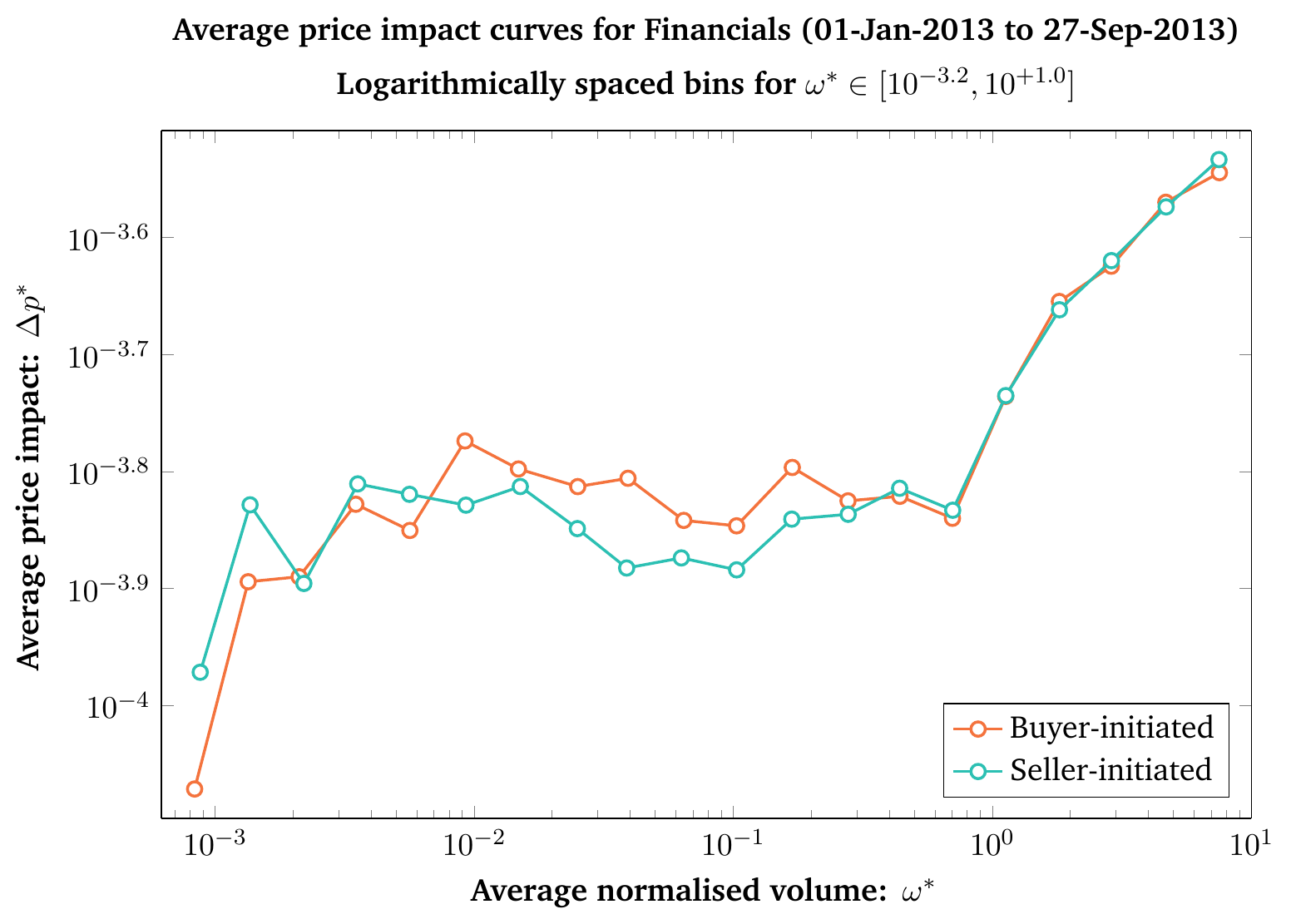}%
}
\hfill
\subfloat[Period after fee structure change.]{%
\includegraphics[width=0.49\textwidth]{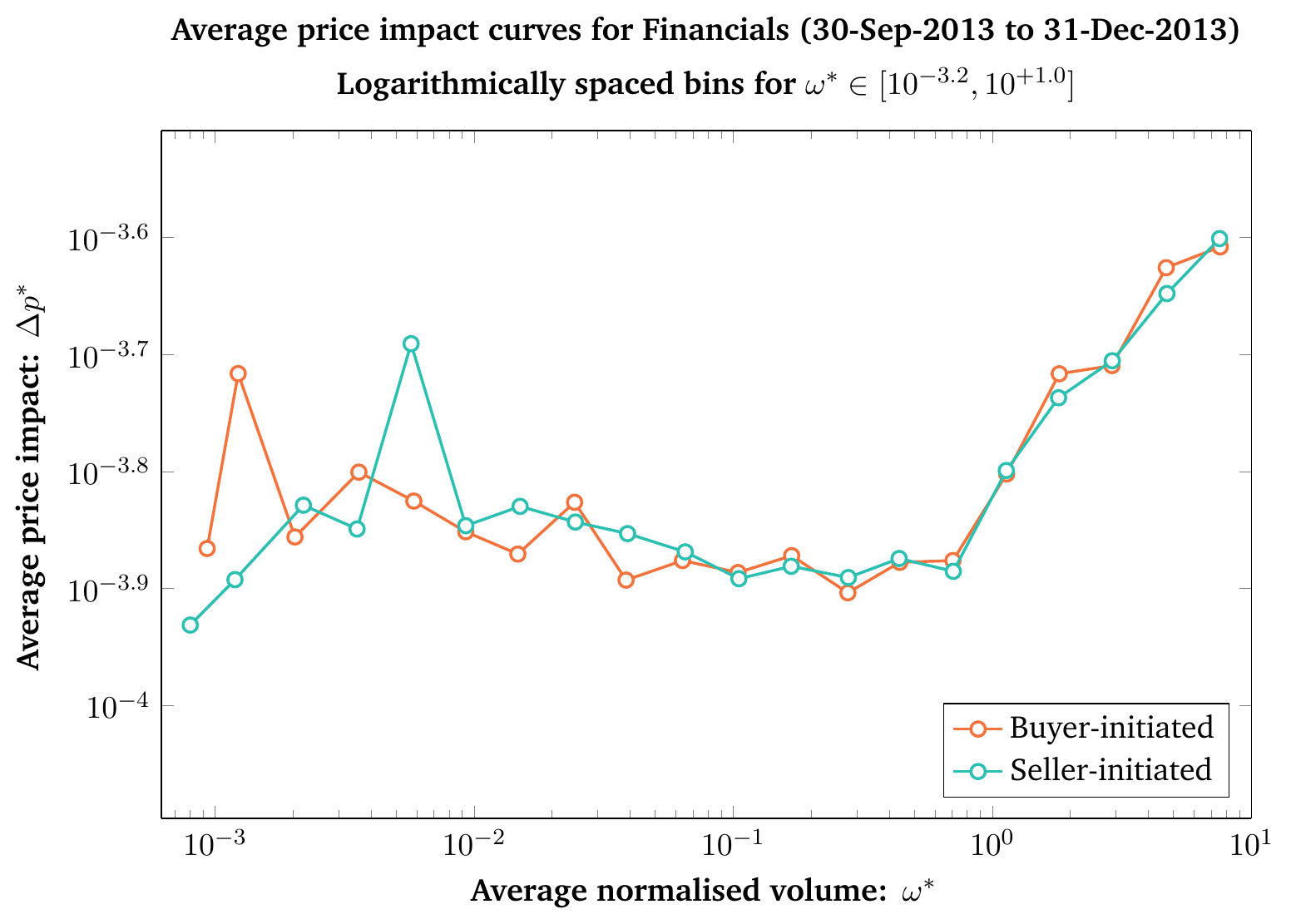}%
}
\caption{Price impact curves for transactions of constituents of the JSE Financials sector (JSE-FINI) for the periods 01-Jan-2013 to  27-Sep-2013 (left) and 30-Sep-2013 to 31-Dec-2013 (right).}\label{PICurveFINI}
\end{figure*}

\begin{figure*}[tbp!]
\centering
\subfloat[Period before fee structure change.]{%
\includegraphics[width=0.49\textwidth]{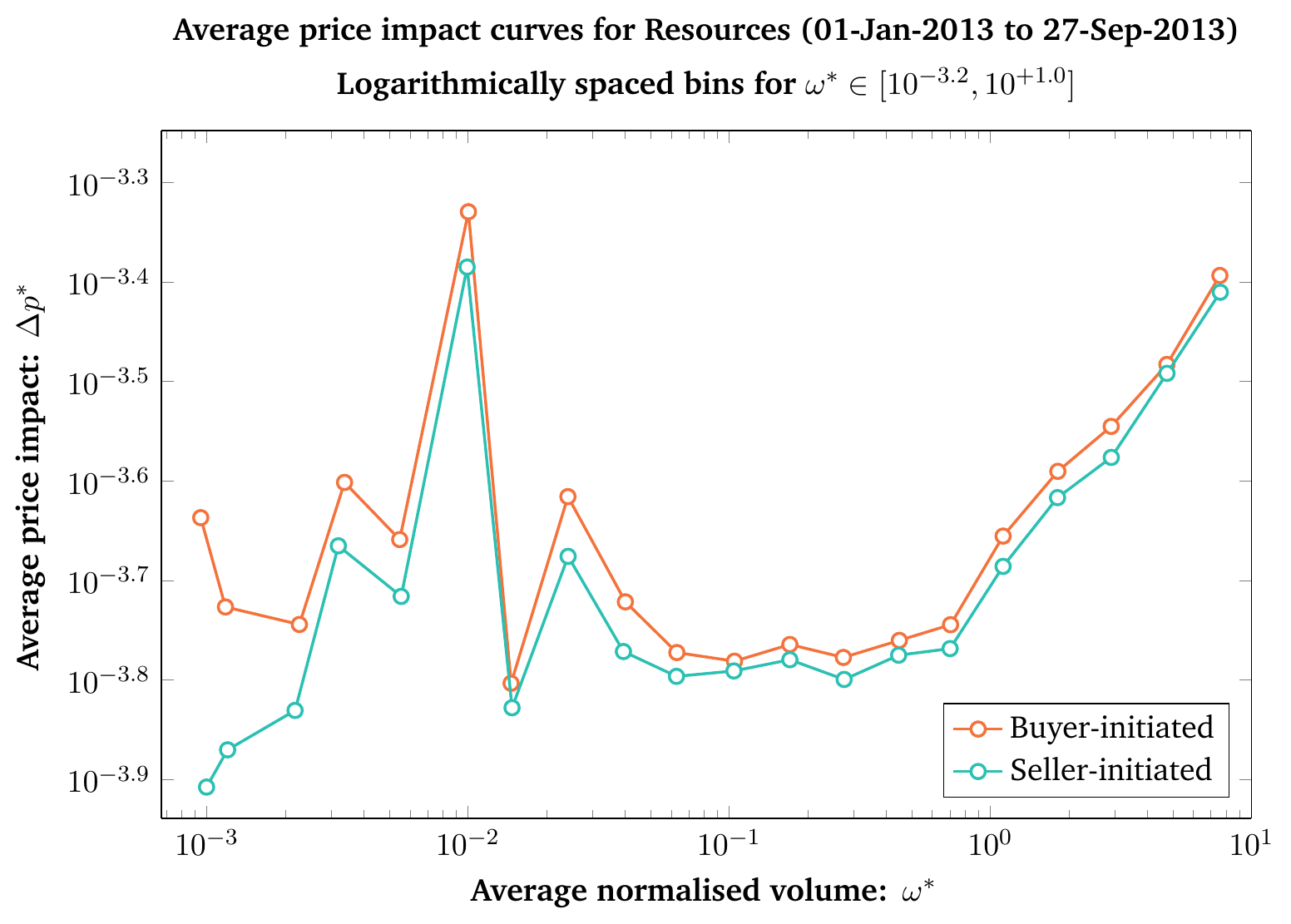}%
}
\hfill
\subfloat[Period after fee structure change.]{%
\includegraphics[width=0.49\textwidth]{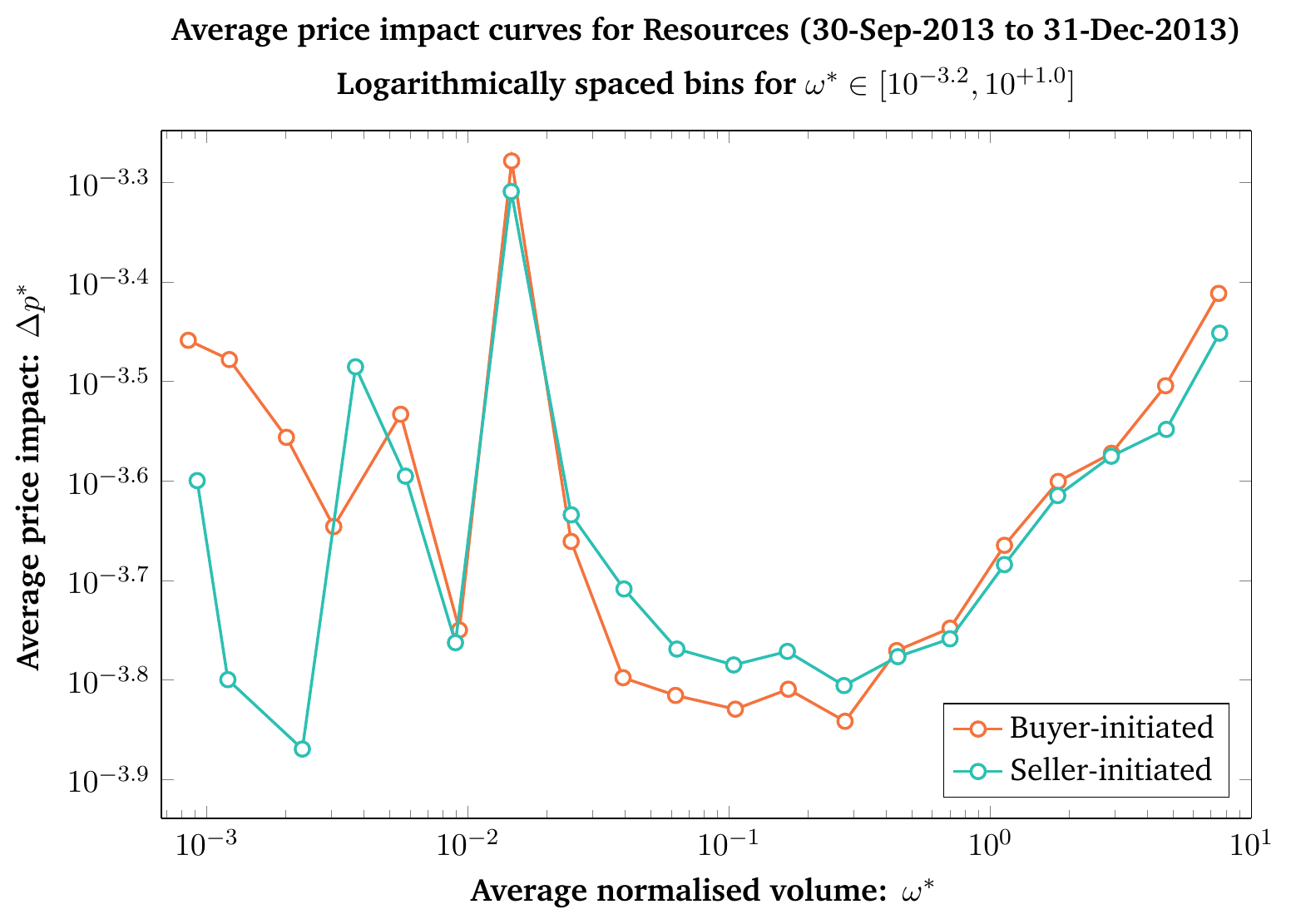}%
}
\caption{Price impact curves for transactions of constituents of the JSE Resources sector (JSE-RESI) for the periods 01-Jan-2013 to  27-Sep-2013 (left) and 30-Sep-2013 to 31-Dec-2013 (right).}\label{PICurveRESI}
\end{figure*}

\begin{figure*}[tbp!]
\centering
\subfloat[Period before fee structure change.]{%
\includegraphics[width=0.49\textwidth]{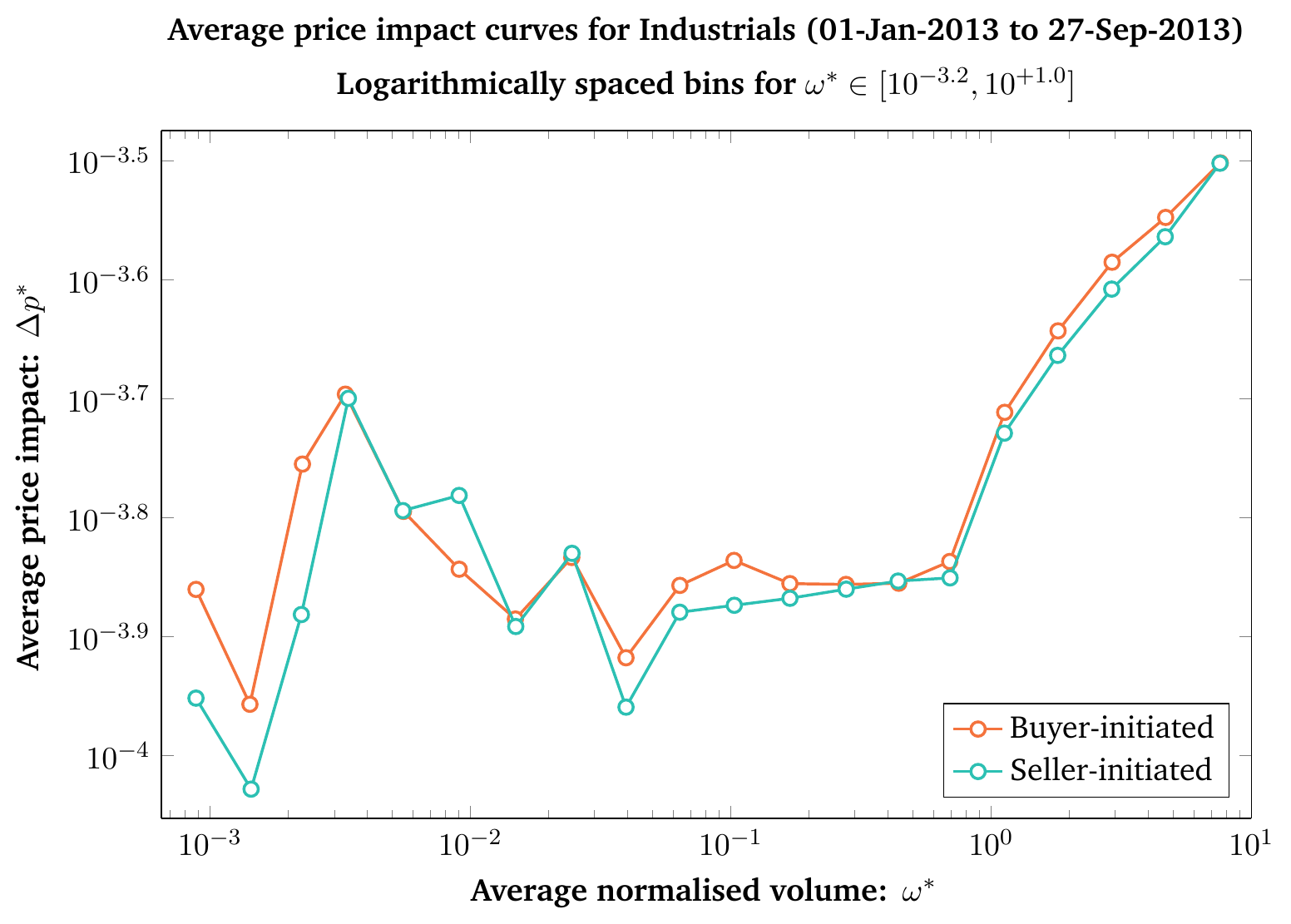}%
}
\hfill
\subfloat[Period after fee structure change.]{%
\includegraphics[width=0.49\textwidth]{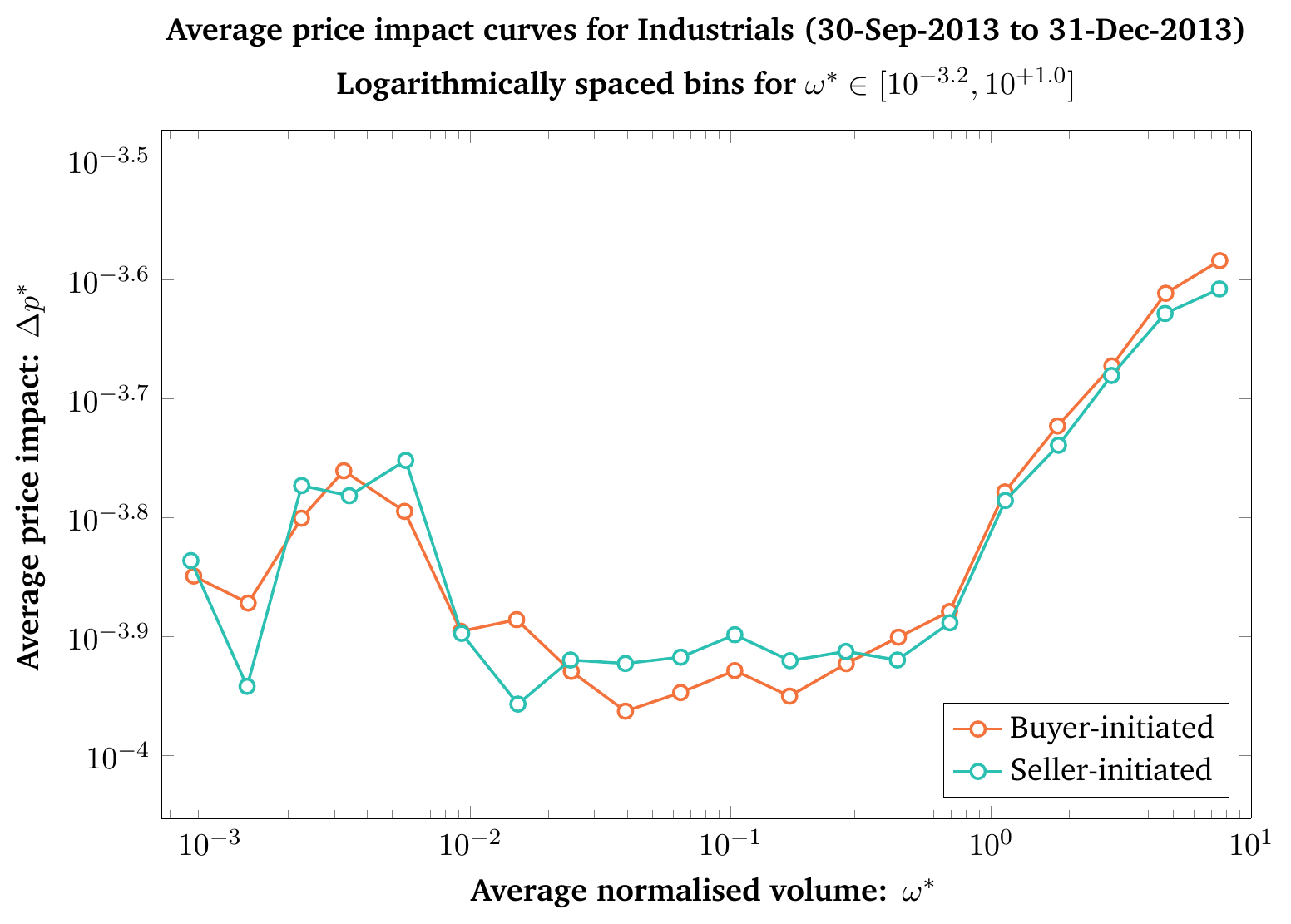}%
}
\caption{Price impact curves for transactions of constituents of the JSE Industrials sector (JSE-INDI) for the periods 01-Jan-2013 to  27-Sep-2013 (left) and 30-Sep-2013 to 31-Dec-2013 (right).}\label{PICurveINDI}
\end{figure*}

\cref{GroupPICurvesBuyer,GroupPICurvesSeller,PICurveFINI,PICurveRESI,PICurveINDI}~plot results regarding the observed anomaly in price impact, viz. the increase in price impact of lower volume trades for a range of stocks after fee restructuring of an electronic market.


\cref{GroupPICurvesBuyer,GroupPICurvesSeller}~provide investigation outcomes of price impact relationships for buyer and seller initiated transactions of the three major sectors in the JSE for the two periods 01-Jan-2013 to 27-Sep-2013 and 01-Oct-2013 to 31-Dec-2013.

By inspection, and taking into account the different ranges for the volume axes, there is a discernible difference in the average price impact of smaller volume trades for stocks in all three sector between the two periods.

\cref{PICurveFINI}~provides closer inspection of the price impact of trades from the Financials sector. Here one can see that the change in the log-midquote price for trades with normalised volume below $10^{-1}$ units have increased from the first to the second period, where average increments for smaller trades increased to values above $10^{-3.8}$ for the most part.

Outcomes for similar comparisons of the two date ranges for the Resources and Industrials sectors are presented in~\cref{PICurveRESI,PICurveINDI}, respectively. For these sectors there also appears to be an increase in price impact for trades with a normalised volume below $10^{-1}$, however the effect is less pronounced compared to Financial stocks.

\ref{app:dscrptvstats}~provides some insight into potential behavioural changes after the fee restructure. One would expect that the removal of the cost floor may result in a higher intensity of low-volume trading activity, since trading schemes which exploit the varying scales of interaction and throughput would not be penalised under the pro-rata fee regime.~%
\cref{EmpiDistBuyerVolumes,EmpiDistSellerVolumes}~confirm an increase in lower-volume trades across all three sectors after the fee restructuring, suggesting a possible change in trader behaviour.~%
\cref{EmpiDistBuyerPriceImpacts,EmpiDistSellerPriceImpacts}~confirm that the distribution of price impacts has shifted to the right following the fee restructure, suggesting that the higher low-volume trade throughput was coupled with an increased associated price impact.~%
\cref{EmpiDistBuyerPrices,EmpiDistSellerPrices}~confirm the overall lower price levels of Financial stocks versus Resource and Industrial stocks. This could explain the observed larger change in low-volume price impact for Financial stocks following the fee restructuring compared to others, since the cost floor was less likely to be a concern for stocks with a high price level prior to the fee restructure, even for small volume trades. The overall higher price impact for Financial stocks versus Resource and Industrial stocks is also consistent with the tick-size versus price argument of Farmer {\em et al.}~\cite{FGLMS2004}, since the tick-size is consistent for stocks across all three sectors, but average price levels of Financials are lower.

Deviation in the expected linear relationship between log-midquote increments and normalised volumes for stocks in our study are consistent with the findings of~\cite{LC2004, Zhou2012, Wilinski2015}, respectively  for stocks listed on the Australian stock exchange between 2001 and 2004,   for 40 constituent stocks in the Shenzhen Stock Exchange Component Index for 2003 and for 6 liquid stocks listed on the FTSE for November 2011 to March 2012. Conditions for the anomaly are proposed in~\cite{Zhou2012}: the condition is derived from proportion of filled trades in small-volume bins under the constraint that the Shenzhen market permits only trade volumes in integer multiples of 100.

One possible explanation could be attributed to a change in order book resiliency before and after the fee restructure~\cite{Large2007,Degryse2005,HH2016}. Hendricks and Harvey conjecture that a decrease in limit order book resiliency for low-volume price-moving trade events may explain the observed increase in price impact~\cite{HH2016}. They study key quote replenishment intensities following low-volume price moving trades. If the observed increase in intensity of low-volume trades is not followed by a commensurate increase in quote replenishment, the less resilient order book may permit higher price impact. This can be studied by investigating the branching ratios of a multivariate Hawkes process calibrated to key limit order book events which are consistent with the price impact study.

\cref{GroupPICurvesBuyer,GroupPICurvesSeller,PICurveFINI,PICurveRESI,PICurveINDI}~reveal an approximate linear relationship between (average) price impact and transaction size (in log scale), for normalised transaction volumes greater than $10^{-0.9}$. To verify the significance of this relationship, and whether it suggests the presence of some invariant law, we perform two investigations: 1) We investigate the existence of a power-law size distribution in the tail of average price impact measured for each sector, and 2) we investigate whether we can find rescaling parameters to reveal a master curve for price impact, following the conjectured power-law dependence between price impact and transaction size, as suggested by~Lillo {\em et al.}~\cite{Lillo2003}. If the distribution of measured avearge price impact quantities fit a power-law for each sector, this would strengthen the argument for rescaling the axes by some liquidity proxy, to reveal an invariant relationship between (average) price impact and normalised transaciton size, referred to as the master curve~\cite{Lillo2003}.

\subsection{Price impact size distribution}

\ref{app:powerlaws}~provides details on power-law size distributions that were fitted to a subset of the average price impact data for buyer and seller initiated transactions of constituents of the three sectors studied.~%
The lower bound on average normalised transaction volume that was used to subset the average price impact data is obtained by inspecting the graphs in~\cref{GroupPICurvesBuyer,GroupPICurvesSeller}.~%
Specifically, power-law size distributions are fit only to those average price impact data that have an average normalised transaction volume exceeding $10^{-0.9}$.

To identify the tail distribution which yields the most significant power-law fit (if present), we employed the maximum-likelihood estimation procedure developed by~\cite{CSN2009}. \cref{tab:PowerLawFitsTable}~shows the estimated exponents, \textit{p}-values and lower bounds of the fitted power-law distributions.~%
Outcomes with \textit{p}-values greater than 0.1 result in a failure to reject the null hypothesis of a power-law distribution fit to the average price impact data for all three sectors.%

The observed statistically significant power law fits are coupled with changes in both scaling exponents ($\alpha$) and minimum values for the fit ($x_{min}$) before and after the fee restructuring.~%
This suggests that the fee restructure may have resulted in a notable change in the typical price impact distribution, however it does not provide us with a direct cause in the context of limit order book dynamics.~%
In addition, variation in the exponents among sectors implies that each sector exhibits distinct price impact dynamics, however rescaling the axes by an appropriate liquidity proxy may reveal an invariant relationship, as was found in~\cite{Lillo2003}.

\subsection{Relationship between price impact and trade volume}

As discussed above, we have chosen to focus on average price impacts for normalised transaction volumes greater than $10^{-0.9}$ to investigate the existence of some invariant relationship via rescaling. Lillo {\em et al.}~\cite{Lillo2003}~used market capitalisation as a liquidity proxy for rescaling the price impact and volume axes, following an observation that the price impact curves were `stacked' according to the associated market capitalisation of the associated stocks. We will consider using average daily value traded as a liquidity proxy to rescale the axes, finding the $\gamma$ and $\delta$ exponents which collapse the curves into a `master curve' for price impact.

\begin{figure*}[tbp!]
\centering
\subfloat[Period before fee structure change.]{%
\includegraphics[width=0.49\textwidth]{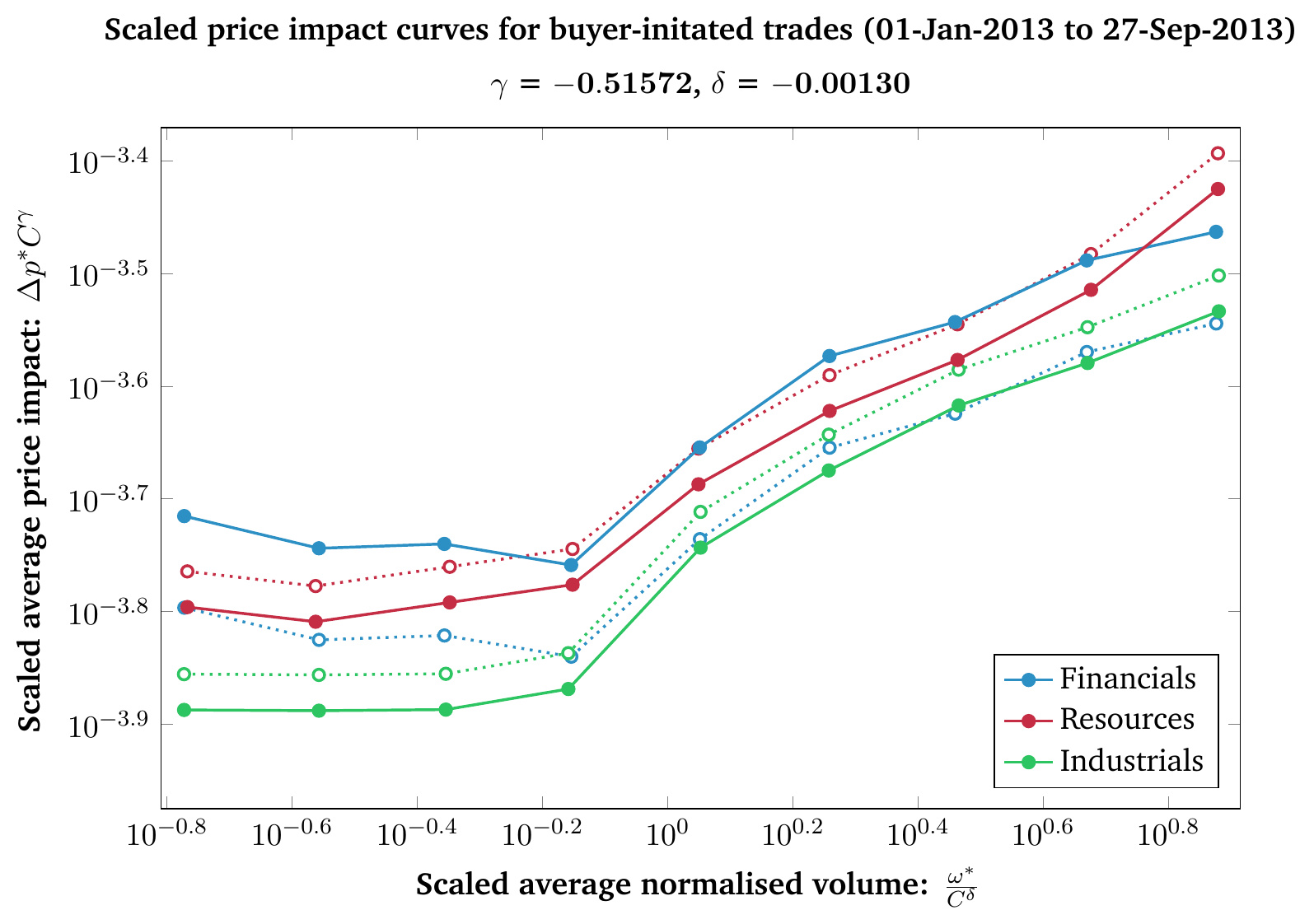}%
}
\hfill
\subfloat[Period after fee structure change.]{%
\includegraphics[width=0.49\textwidth]{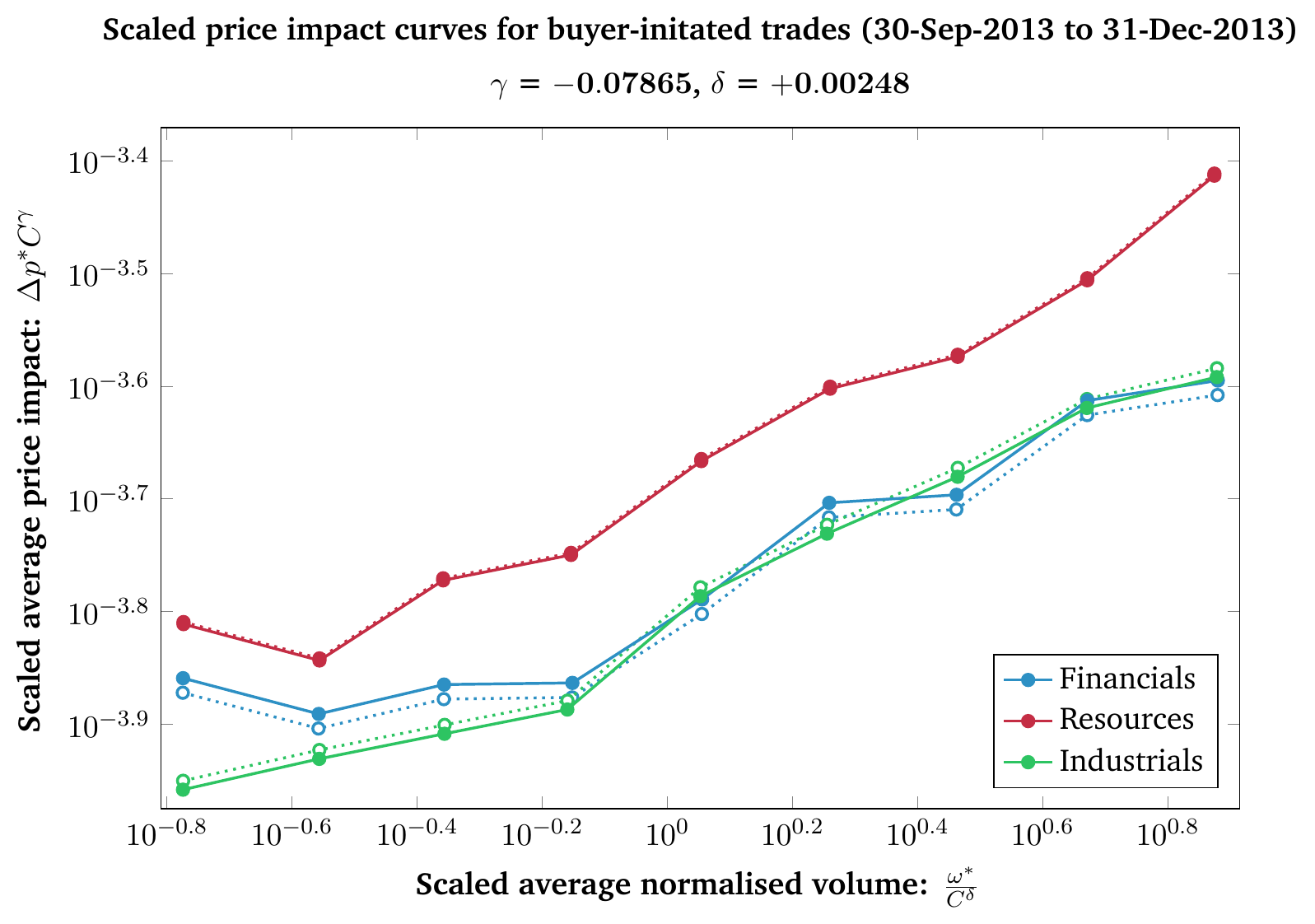}%
}
\caption{%
Collapse of price impact curves for buyer initiated transactions of constituents of the Financials (JSE-FINI), Resources (JSE-RESI) and Industrials (JSE-INDI) sectors for the periods 01-Jan-2013 to  27-Sep-2013 (left) and 30-Sep-2013 to 31-Dec-2013 (right).~%
Price impact data with normalised transaction volumes strictly less than $10^{-0.9}$ are excluded from the data used to collapse the curves.~%
The three dashed lines with unfilled dots are the price impact curves for the three sectors prior to scaling by the liquidity adjustment parameter $C$. The three solid lines with filled dots are the shifted price impact curves obtained when scaling the average price impact data pairs $\left(\omega^{\ast},\Delta p^{\ast}\right)$ by the parameters $\gamma$ and $\delta$ estimated when collapsing the three sector curves into a single best fit curve with the liquidity adjustment $C$.}\label{MasterCurveBuyer}
\end{figure*}

\begin{figure*}[tbp!]
\centering
\subfloat[Period before fee structure change.]{%
\includegraphics[width=0.49\textwidth]{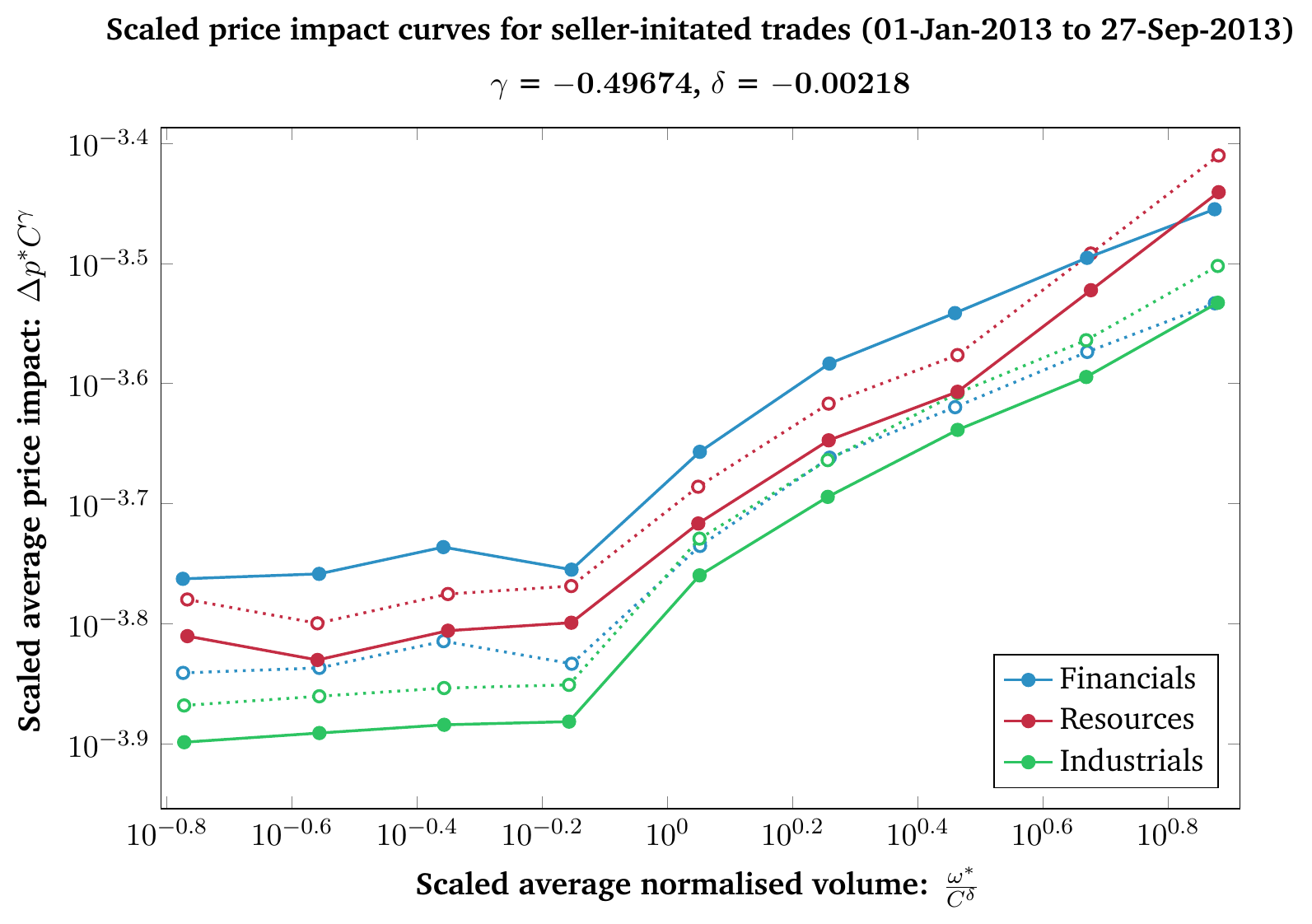}%
}
\hfill
\subfloat[Period after fee structure change.]{%
\includegraphics[width=0.49\textwidth]{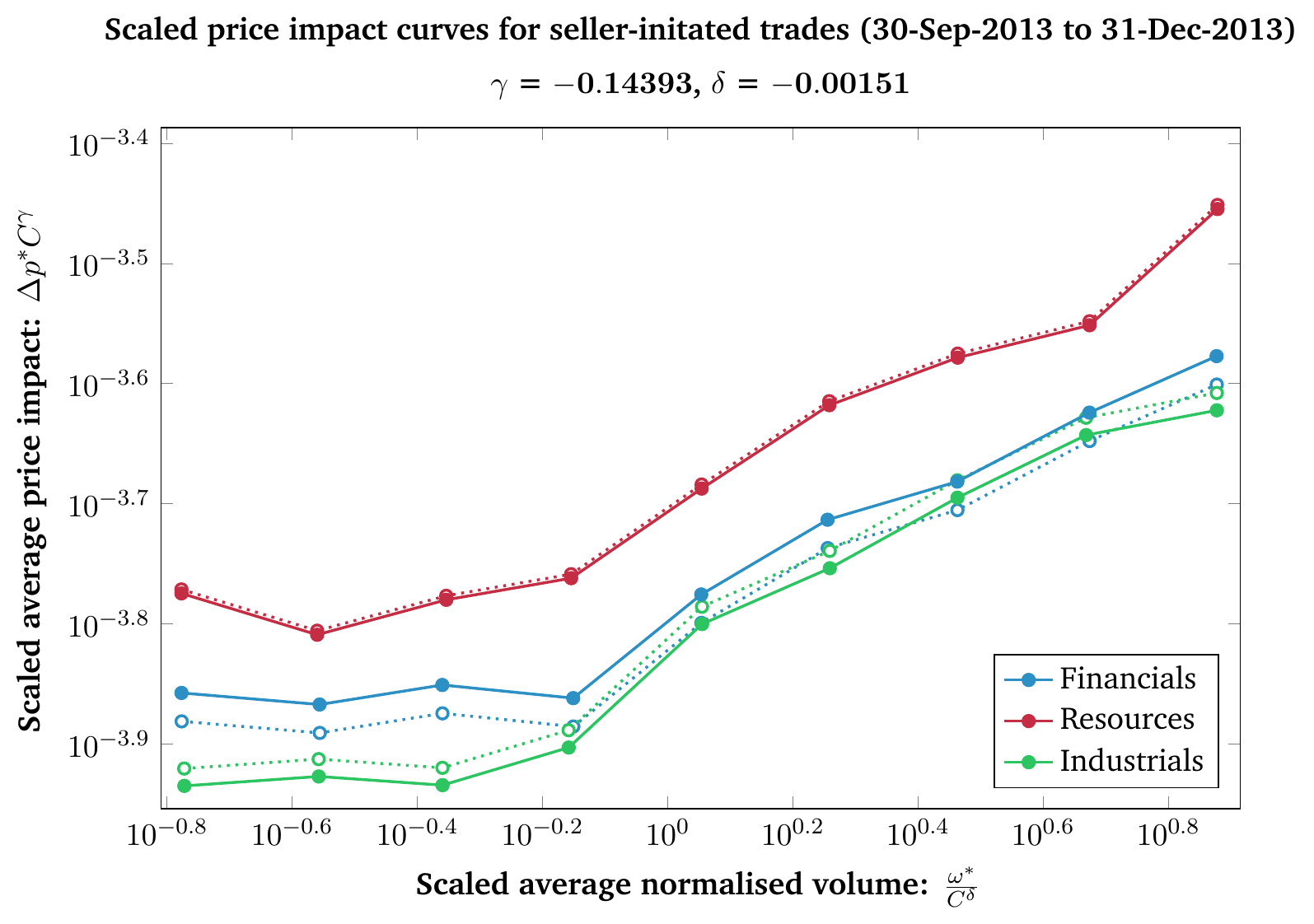}%
}
\caption{%
Collapse of price impact curves for seller initiated transactions of constituents of the Financials (JSE-FINI), Resources (JSE-RESI) and Industrials (JSE-INDI) sectors for the periods 01-Jan-2013 to  27-Sep-2013 (left) and 30-Sep-2013 to 31-Dec-2013 (right).~%
Price impact data with normalised transaction volumes strictly less than $10^{-0.9}$ are excluded from the data used to collapse the curves.~%
The three dashed lines with unfilled dots are the price impact curves for the three sectors prior to scaling by the liquidity adjustment parameter $C$. The three solid lines with filled dots are the shifted price impact curves obtained when scaling the average price impact data pairs $\left(\omega^{\ast},\Delta p^{\ast}\right)$ by the parameters $\gamma$ and $\delta$ estimated when collapsing the three sector curves into a single best fit curve with the liquidity adjustment $C$.}\label{MasterCurveSeller}
\end{figure*}

\cref{MasterCurveBuyer,MasterCurveSeller}~plot results regarding the collapse of the average price impact curves for data with normalised transaction volumes greater than $10^{-0.9}$. To perform the collapse the method of~Lillo {\em et al.}~\cite{Lillo2003}~was used where the parameters $\gamma$ and $\delta$ are chosen in order to best collapse all data onto a single impact curve. To do this, the average normalised volume axis (the $x$-axis) is first divided into $N_{bins}$. For each bin the mean ($\mu^{(k)}$) and standard deviation ($\sigma^{(k)}$) are computed for both the renormalised volume ($x \rightarrow {\omega^{\ast}}/{C^{\delta}}$) and renormalised price impact ($y \rightarrow \Delta p^{\ast} C^{\gamma}$) components. The parameters $\gamma$ and $\delta$ are then estimated such that they minimize the average two-dimensional variance:
\begin{align*}
\epsilon = \frac{1}{N_{b}}\sum_{k=1}^{N_{b}}%
\left[%
\left(\frac{\sigma_{x}^{(k)}}{\mu_{x}^{(k)}}\right)^2 + \left(\frac{\sigma_{y}^{(k)}}{\mu_{y}^{(k)}}\right)^2%
\right]
\end{align*}


~

By calibrating a master curve for the relationship between price impact and trade volume, we have effectively provided a means for recovering the underlying price impact curve for a particular sector, given only its liquidity proxy and scaling exponents ($\delta$ and $\gamma$). This is of significant importance for practitioners, as it provides a quick and efficient means to approximate price impact curves without extensive processing of tick data, however periodic recalibration of the scaling exponents will be required to ensure relevant regimes are incorporated.

\section{Conclusion}\label{conclusion}

We investigated the nature of price impact of electronic trading of share prices in an emerging economy, with attention to change in market microstructure through transaction fee restructuring. For the three major sectors studied we confirmed:
\begin{enumerate}
   \item the existence of a  power-law relationship between price increment and transaction size and
   \item that the relationship was not changed significantly by reduction of transaction costs for the period under investigation.
\end{enumerate}
We detected a discernible anomaly in the price impact of small trades for stocks across all three major sectors, especially Financials, where it was found that measured price shifts were greater than expected for the smaller volumes traded.

The deviation  has significant implications for risk-management: a decrease in direct transaction costs resulted in an unexpected increase in indirect costs through price impact.

Assuming that price impact models are incorporated into optimal trading strategies for dynamic hedging, this would have translated to an increase in unexpected hedging costs for derivative instruments on a range of stocks.

The periodic calibration of a so-called price impact master curve to certain groupings of stocks is useful because it provides the practitioner with a quick and efficient way of approximating a single impact curve for a new grouping of stocks using only the liquidity proxy of that grouping and the scaling exponents ($\delta$ and $\gamma$). The practitioner is thus able to approximate the relationship between transaction volume and price impact without the need to process large quantities of tick data.%

The investigation contributes to the general analysis of electronic markets as an observational science, since measures of effects of change in regulation may only be inferred from past behaviour and not from repeated experiment.

Further investigation should include closer analysis of the order books of stocks studied here, using more in depth market-quality measures,  in order to uncover a comprehensive explanation for the observed anomaly.

Hendricks and Harvey~\cite{HH2016}~have suggested a possible explanation for the observed increase in low-volume price impact using an argument of decreased order book resiliency, however these conclusions still need to be verified.

\section*{Acknowledgement}
This work is based on the research supported in part by the National Research Foundation of South Africa (Grant number 87830). The conclusions herein are due to the authors and the NRF accepts no liability in this regard.

\bibliography{BibFile}

\begin{thebibliography}{29}
\expandafter\ifx\csname natexlab\endcsname\relax\def\natexlab#1{#1}\fi
\providecommand{\url}[1]{\texttt{#1}}
\providecommand{\href}[2]{#2}
\providecommand{\path}[1]{#1}
\providecommand{\DOIprefix}{doi:}
\providecommand{\ArXivprefix}{arXiv:}
\providecommand{\URLprefix}{URL: }
\providecommand{\Pubmedprefix}{pmid:}
\providecommand{\doi}[1]{\href{http://dx.doi.org/#1}{\path{#1}}}
\providecommand{\Pubmed}[1]{\href{pmid:#1}{\path{#1}}}
\providecommand{\bibinfo}[2]{#2}
\ifx\xfnm\relax \def\xfnm[#1]{\unskip,\space#1}\fi
\bibitem[{Bouchaud et~al.(2004)Bouchaud, Gefen, Potters and Wyart}]{BGPW2004}
\bibinfo{author}{Bouchaud, J.P.}, \bibinfo{author}{Gefen, Y.},
  \bibinfo{author}{Potters, M.}, \bibinfo{author}{Wyart, M.},
  \bibinfo{year}{2004}.
\newblock \bibinfo{title}{{Fluctuations and response in financial markets: the
  subtle nature of `random' price changes}}.
\newblock \bibinfo{journal}{Quantitative Finance} \bibinfo{volume}{4},
  \bibinfo{pages}{176--190}.
\newblock \DOIprefix\doi{10.1080/14697680400000022}.
\bibitem[{Clauset et~al.(2009)Clauset, Shalizi and Newman}]{CSN2009}
\bibinfo{author}{Clauset, A.}, \bibinfo{author}{Shalizi, C.R.},
  \bibinfo{author}{Newman, M.E.J.}, \bibinfo{year}{2009}.
\newblock \bibinfo{title}{{Power law distributions in empirical data}}.
\newblock \bibinfo{journal}{SIAM Rev.} \bibinfo{volume}{51},
  \bibinfo{pages}{661--703}.
\newblock \DOIprefix\doi{10.1137/070710111}.
\bibitem[{Correia and Uliana(2004)}]{CU2004}
\bibinfo{author}{Correia, C.}, \bibinfo{author}{Uliana, E.},
  \bibinfo{year}{2004}.
\newblock \bibinfo{title}{{Market segmentation and the cost of equity of
  companies listed on the Johannesburg Stock Exchange}}.
\newblock \bibinfo{journal}{South African Journal of Accounting Research}
  \bibinfo{volume}{18}, \bibinfo{pages}{65--86}.
\newblock \DOIprefix\doi{10.1080/10291954.2004.11435109}.
\bibitem[{Degryse et~al.(2005)Degryse, Jong, Ravenswaaij and
  Wuyts}]{Degryse2005}
\bibinfo{author}{Degryse, H.}, \bibinfo{author}{Jong, F.D.},
  \bibinfo{author}{Ravenswaaij, M.V.}, \bibinfo{author}{Wuyts, G.},
  \bibinfo{year}{2005}.
\newblock \bibinfo{title}{{Aggressive Orders and the Resiliency of a Limit
  Order Market}}.
\newblock \bibinfo{journal}{Review of Finance} \bibinfo{volume}{9},
  \bibinfo{pages}{201--242}.
\newblock \DOIprefix\doi{10.1007/s10679-005-7590-6}.
\bibitem[{{Du~Preez}(2015)}]{Dupreez}
\bibinfo{author}{{Du~Preez}, B.S.}, \bibinfo{year}{2015}.
\newblock \bibinfo{title}{{JSE Market Micro-Structure}}.
\newblock Master's thesis. University of the Witwatersrand.
  \bibinfo{address}{Johannesburg}.
\bibitem[{Farmer et~al.(2004)Farmer, Gillemot, Lillo, Mike and Sen}]{FGLMS2004}
\bibinfo{author}{Farmer, J.D.}, \bibinfo{author}{Gillemot, L.},
  \bibinfo{author}{Lillo, F.}, \bibinfo{author}{Mike, S.},
  \bibinfo{author}{Sen, A.}, \bibinfo{year}{2004}.
\newblock \bibinfo{title}{{What really causes large price changes?}}
\newblock \bibinfo{journal}{Quantitative Finance} \bibinfo{volume}{4},
  \bibinfo{pages}{383--397}.
\newblock \DOIprefix\doi{10.1080/14697680400008627}.
\bibitem[{Farmer and Lillo(2004)}]{Farmer2004}
\bibinfo{author}{Farmer, J.D.}, \bibinfo{author}{Lillo, F.},
  \bibinfo{year}{2004}.
\newblock \bibinfo{title}{{On the origin of power-law tails in price
  fluctuations}}.
\newblock \bibinfo{journal}{Quantitative Finance} \bibinfo{volume}{4},
  \bibinfo{pages}{7--11}.
\newblock \DOIprefix\doi{10.1088/1469-7688/4/1/C01}.
\bibitem[{Gatheral(2010)}]{Gatheral2009}
\bibinfo{author}{Gatheral, J.}, \bibinfo{year}{2010}.
\newblock \bibinfo{title}{{No-dynamic-arbitrage and market impact}}.
\newblock \bibinfo{journal}{Quantitative Finance} \bibinfo{volume}{10},
  \bibinfo{pages}{749--759}.
\newblock \DOIprefix\doi{10.1080/14697680903373692}.
\bibitem[{Hendricks and Harvey(2016)}]{HH2016}
\bibinfo{author}{Hendricks, D.}, \bibinfo{author}{Harvey, M.},
  \bibinfo{year}{2016}.
\newblock \bibinfo{title}{{Reconciling order book resiliency and price
  impact}}.
\bibitem[{Hendricks and Wilcox(2014)}]{Hendricks2014}
\bibinfo{author}{Hendricks, D.}, \bibinfo{author}{Wilcox, D.},
  \bibinfo{year}{2014}.
\newblock \bibinfo{title}{{A reinforcement learning extension to the
  Almgren-Chriss framework for optimal trade execution}}, in:
  \bibinfo{booktitle}{{2014 IEEE Conference on Computational Intelligence for
  Financial Engineering and Economics (CIFEr)}}, \bibinfo{publisher}{IEEE}. pp.
  \bibinfo{pages}{457--464}.
\newblock \DOIprefix\doi{10.1109/CIFEr.2014.6924109}.
\bibitem[{Iori et~al.(2003)Iori, Daniels, Farmer, Gillemot, Krishnamurthy and
  Smith}]{Iori2003}
\bibinfo{author}{Iori, G.}, \bibinfo{author}{Daniels, M.G.},
  \bibinfo{author}{Farmer, J.D.}, \bibinfo{author}{Gillemot, L.},
  \bibinfo{author}{Krishnamurthy, S.}, \bibinfo{author}{Smith, E.},
  \bibinfo{year}{2003}.
\newblock \bibinfo{title}{{An analysis of price impact function in order-driven
  markets}}.
\newblock \bibinfo{journal}{Physica A: Statistical Mechanics and its
  Applications} \bibinfo{volume}{324}, \bibinfo{pages}{146--151}.
\newblock \DOIprefix\doi{10.1016/S0378-4371(02)01888-5}.
  \bibinfo{note}{proceedings of the International Econophysics Conference}.
\bibitem[{{Johannesburg Stock Exchange}(2014a)}]{JSE:ColocationGoesLive2014}
\bibinfo{author}{{Johannesburg Stock Exchange}}, \bibinfo{year}{2014}a.
\newblock \bibinfo{title}{{JSE Colocation Services Go Live 12 May 2014}}.
\newblock
  \bibinfo{howpublished}{\url{https://www.jse.co.za/content/JSEHotlinesItems/JSE\%20Service\%20Hotline\%209014\%20JSE\%20Colocation\%20Service\%20Go\%20live\%2012\%20May\%202014.pdf}}.
\newblock \bibinfo{note}{Online; accessed February 2016}.
\bibitem[{{Johannesburg Stock Exchange}(2014b)}]{JSE:ColocationBrochure2014}
\bibinfo{author}{{Johannesburg Stock Exchange}}, \bibinfo{year}{2014}b.
\newblock \bibinfo{title}{{The lowest-latency connection to JSE markets:
  Colocation}}.
\newblock
  \bibinfo{howpublished}{\url{https://www.jse.co.za/content/JSETechnologyDocumentItems/3.\%20JSE\%20Colocation\%20Brochure\%202015.pdf}}.
\newblock \bibinfo{note}{Online; accessed February 2016}.
\bibitem[{{JSE Market Notices and
  Circulars}(2012a)}]{JSE:EquityMarketPriceList2013}
\bibinfo{author}{{JSE Market Notices and Circulars}}, \bibinfo{year}{2012}a.
\newblock \bibinfo{title}{{Equity Market Price List 2013}}.
\newblock
  \bibinfo{howpublished}{\url{https://www.jse.co.za/content/JSENoticesandCircularsItems/Equity\%20Markets/2012/20121130-098B-.pdf}}.
\newblock \bibinfo{note}{Online; accessed February 2016}.
\bibitem[{{JSE Market Notices and
  Circulars}(2012b)}]{JSE:MilleniumITGoLiveNotice2012}
\bibinfo{author}{{JSE Market Notices and Circulars}}, \bibinfo{year}{2012}b.
\newblock \bibinfo{title}{{New Equity Market Trading and Information Solution
  and New SENS System - Go Live Readiness Confirmed for Monday 2 July 2012}}.
\newblock
  \bibinfo{howpublished}{\url{https://www.jse.co.za/content/JSENoticesandCircularsItems/Equity\%20Markets/2012/20120629-062.pdf}}.
\newblock \bibinfo{note}{Online; accessed February 2016}.
\bibitem[{{JSE Market Notices and
  Circulars}(2013)}]{JSE:EquityMarketBillingModelChangeNotice2013}
\bibinfo{author}{{JSE Market Notices and Circulars}}, \bibinfo{year}{2013}.
\newblock \bibinfo{title}{{JSE Equity Market Transaction Billing Model
  Methodology Change Notice, JSE Market Notice No. 136}}.
\newblock
  \bibinfo{howpublished}{\url{https://www.jse.co.za/content/JSENoticesandCircularsItems/Equity\%20Markets/2013/2013_136.pdf}}.
\newblock \bibinfo{note}{Online; accessed February 2016}.
\bibitem[{{JSE Market Notices and
  Circulars}(2014)}]{JSE:EquityMarketPriceList2014v1a}
\bibinfo{author}{{JSE Market Notices and Circulars}}, \bibinfo{year}{2014}.
\newblock \bibinfo{title}{{Equity Market Price List 2014 v1.1}}.
\newblock
  \bibinfo{howpublished}{\url{https://www.jse.co.za/content/JSENoticesandCircularsItems/Equity\%20Markets/2014/218B.pdf}}.
\newblock \bibinfo{note}{Online; accessed February 2016}.
\bibitem[{Large(2007)}]{Large2007}
\bibinfo{author}{Large, J.}, \bibinfo{year}{2007}.
\newblock \bibinfo{title}{{Measuring the resiliency of an electronic limit
  order book}}.
\newblock \bibinfo{journal}{Journal of Financial Markets} \bibinfo{volume}{10},
  \bibinfo{pages}{1--25}.
\newblock \DOIprefix\doi{10.1016/j.finmar.2006.09.001}.
\bibitem[{Lee and Ready(1991)}]{LeeReady1991}
\bibinfo{author}{Lee, C.M.C.}, \bibinfo{author}{Ready, M.J.},
  \bibinfo{year}{1991}.
\newblock \bibinfo{title}{{Inferring Trade Direction from Intraday Data}}.
\newblock \bibinfo{journal}{Journal of Finance} \bibinfo{volume}{46},
  \bibinfo{pages}{733--736}.
\newblock \DOIprefix\doi{10.1111/j.1540-6261.1991.tb02683.x}.
\bibitem[{Lillo et~al.(2003)Lillo, Farmer and Mantegna}]{Lillo2003}
\bibinfo{author}{Lillo, F.}, \bibinfo{author}{Farmer, J.D.},
  \bibinfo{author}{Mantegna, R.N.}, \bibinfo{year}{2003}.
\newblock \bibinfo{title}{{Econophysics: Master curve for price-impact
  function}}.
\newblock \bibinfo{journal}{Nature} \bibinfo{volume}{421},
  \bibinfo{pages}{129--130}.
\newblock \DOIprefix\doi{10.1038/421129a}.
\bibitem[{Lim and Coggins(2005)}]{LC2004}
\bibinfo{author}{Lim, M.}, \bibinfo{author}{Coggins, R.}, \bibinfo{year}{2005}.
\newblock \bibinfo{title}{{The immediate price impact of trades on the
  Australian stock exchange}}.
\newblock \bibinfo{journal}{Quantitative Finance} \bibinfo{volume}{5},
  \bibinfo{pages}{365--377}.
\newblock \DOIprefix\doi{10.1080/14697680500151400}.
\bibitem[{Malinova and Park(2015)}]{Malinova2011}
\bibinfo{author}{Malinova, K.}, \bibinfo{author}{Park, A.},
  \bibinfo{year}{2015}.
\newblock \bibinfo{title}{{Subsidizing Liquidity: The Impact of Make/Take Fees
  on Market Quality}}.
\newblock \bibinfo{journal}{The Journal of Finance} \bibinfo{volume}{70},
  \bibinfo{pages}{509--536}.
\newblock \DOIprefix\doi{10.1111/jofi.12230}.
\bibitem[{Moro et~al.(2009)Moro, Vicente, Moyano, Gerig, Farmer, Vaglica, Lillo
  and Mantegna}]{Moro2009}
\bibinfo{author}{Moro, E.}, \bibinfo{author}{Vicente, J.},
  \bibinfo{author}{Moyano, L.G.}, \bibinfo{author}{Gerig, A.},
  \bibinfo{author}{Farmer, J.D.}, \bibinfo{author}{Vaglica, G.},
  \bibinfo{author}{Lillo, F.}, \bibinfo{author}{Mantegna, R.N.},
  \bibinfo{year}{2009}.
\newblock \bibinfo{title}{{Market impact and trading profile of hidden orders
  in stock markets}}.
\newblock \bibinfo{journal}{Phys. Rev. E} \bibinfo{volume}{80},
  \bibinfo{pages}{066102}.
\newblock \DOIprefix\doi{10.1103/PhysRevE.80.066102}.
\bibitem[{Plerou et~al.(2004)Plerou, Stanley, Gabaix and
  Gopikrishnan}]{Plerou2004}
\bibinfo{author}{Plerou, V.}, \bibinfo{author}{Stanley, H.E.},
  \bibinfo{author}{Gabaix, X.}, \bibinfo{author}{Gopikrishnan, P.},
  \bibinfo{year}{2004}.
\newblock \bibinfo{title}{{On the origin of power-law fluctuations in stock
  prices}}.
\newblock \bibinfo{journal}{Quantitative Finance} \bibinfo{volume}{4},
  \bibinfo{pages}{11--15}.
\newblock \DOIprefix\doi{10.1088/1469-7688/4/1/C02}.
\bibitem[{Potters and Bouchaud(2003)}]{Potters2003}
\bibinfo{author}{Potters, M.}, \bibinfo{author}{Bouchaud, J.P.},
  \bibinfo{year}{2003}.
\newblock \bibinfo{title}{{More statistical properties of order books and price
  impact}}.
\newblock \bibinfo{journal}{Physica A: Statistical Mechanics and its
  Applications} \bibinfo{volume}{324}, \bibinfo{pages}{133--140}.
\newblock \DOIprefix\doi{10.1016/S0378-4371(02)01896-4}.
  \bibinfo{note}{proceedings of the International Econophysics Conference}.
\bibitem[{Wilinski et~al.(2015)Wilinski, Cui, Brabazon and
  Hamill}]{Wilinski2015}
\bibinfo{author}{Wilinski, M.}, \bibinfo{author}{Cui, W.},
  \bibinfo{author}{Brabazon, A.}, \bibinfo{author}{Hamill, P.},
  \bibinfo{year}{2015}.
\newblock \bibinfo{title}{{An analysis of price impact functions of individual
  trades on the London stock exchange}}.
\newblock \bibinfo{journal}{Quantitative Finance} \bibinfo{volume}{15},
  \bibinfo{pages}{1727--1735}.
\newblock \DOIprefix\doi{10.1080/14697688.2015.1071077}.
\bibitem[{Yang(2011)}]{Yang2011}
\bibinfo{author}{Yang, J.W.}, \bibinfo{year}{2011}.
\newblock \bibinfo{title}{{Transaction duration and asymmetric price impact of
  trades—Evidence from Australia}}.
\newblock \bibinfo{journal}{Journal of Empirical Finance} \bibinfo{volume}{18},
  \bibinfo{pages}{91--102}.
\newblock \DOIprefix\doi{10.1016/j.jempfin.2010.09.001}.
\bibitem[{Zawadowski et~al.(2004)Zawadowski, Kertész and
  Andor}]{Zawadowski2004}
\bibinfo{author}{Zawadowski, A.G.}, \bibinfo{author}{Kertész, J.},
  \bibinfo{author}{Andor, G.}, \bibinfo{year}{2004}.
\newblock \bibinfo{title}{{Large price changes on small scales}}.
\newblock \bibinfo{journal}{Physica A: Statistical Mechanics and its
  Applications} \bibinfo{volume}{344}, \bibinfo{pages}{221--226}.
\newblock \DOIprefix\doi{10.1016/j.physa.2004.06.121}.
  \bibinfo{note}{applications of Physics in Financial Analysis 4 (APFA4)}.
\bibitem[{Zhou(2012)}]{Zhou2012}
\bibinfo{author}{Zhou, W.X.}, \bibinfo{year}{2012}.
\newblock \bibinfo{title}{{Universal price impact functions of individual
  trades in an order-driven market}}.
\newblock \bibinfo{journal}{Quantitative Finance} \bibinfo{volume}{12},
  \bibinfo{pages}{1253--1263}.
\newblock \DOIprefix\doi{10.1080/14697688.2010.504733}.

\end{thebibliography}
\label{sec:references}

\appendix%

\section{Empirically estimated distributions for daily averages}\label{app:dscrptvstats}

\cref{%
EmpiDistBuyerVolumes,EmpiDistSellerVolumes,%
EmpiDistBuyerPriceImpacts,EmpiDistSellerPriceImpacts,%
EmpiDistBuyerPrices,EmpiDistSellerPrices,%
}~%
plot empirically estimated distributions of the daily averages of the volume, price impact and price for transactions of constituents of the Financials (JSE-FINI), Resources (JSE-RESI) and Industrials (JSE-INDI) sectors for the periods 01-Jan-2013 to  27-Sep-2013 (left) and 30-Sep-2013 to 31-Dec-2013 (right). The plots on the left (right) show the distribution of the daily average computed over the 186 (64) trading days before (after) the fee structure change. Each distribution is estimated by normalising the histogram bin counts of $\left\lceil N \right\rceil$ equally spaced bins where $N$ is either 186 or 64 depending on the period. To aid with visualisation the x-axis is plotted on the logarithmic scale.~%
\begin{figure*}[tbp!]
\centering
\subfloat[Period before fee structure change.]{%
\includegraphics[width=0.49\textwidth]{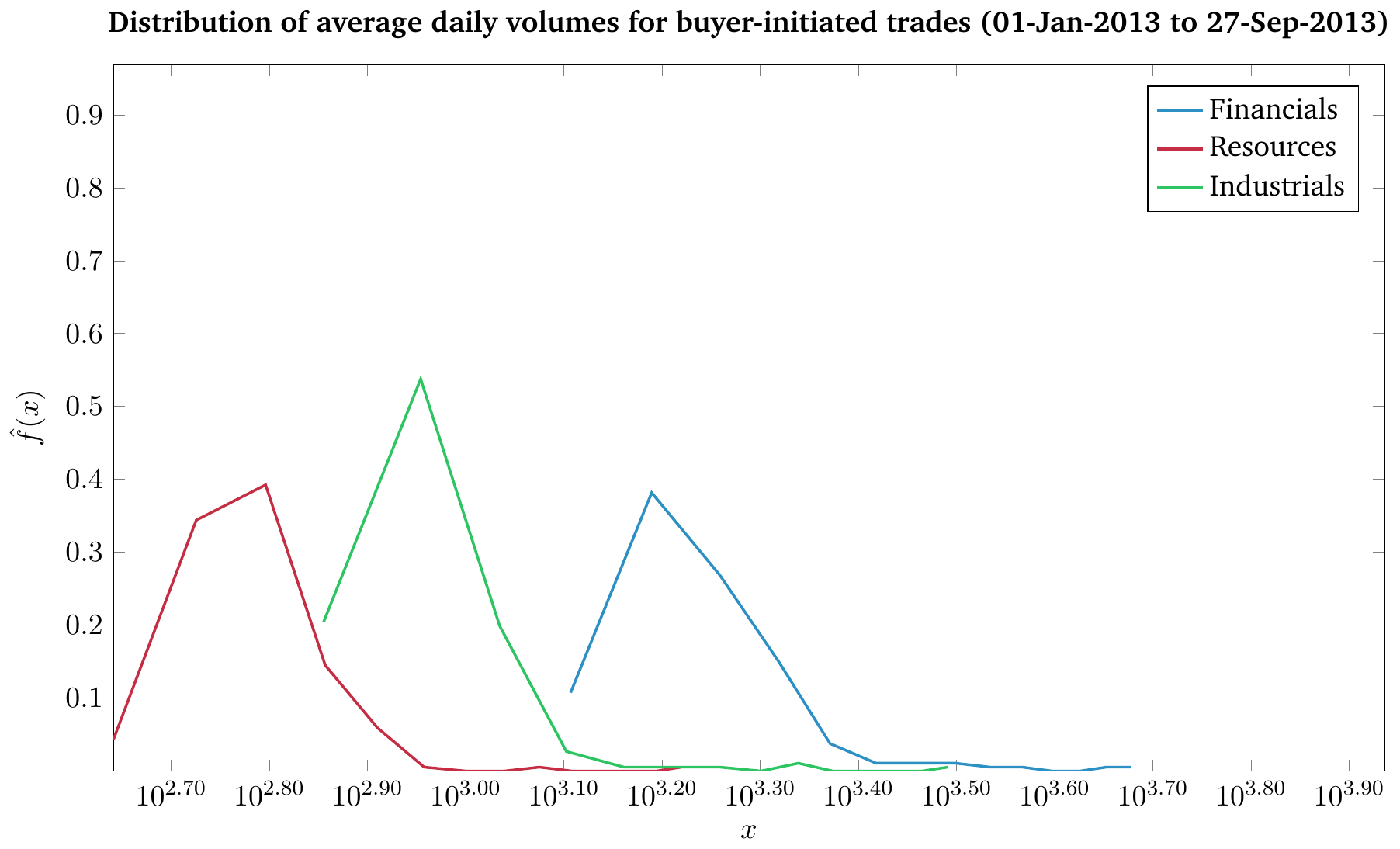}%
}
\hfill
\subfloat[Period after fee structure change.]{%
\includegraphics[width=0.49\textwidth]{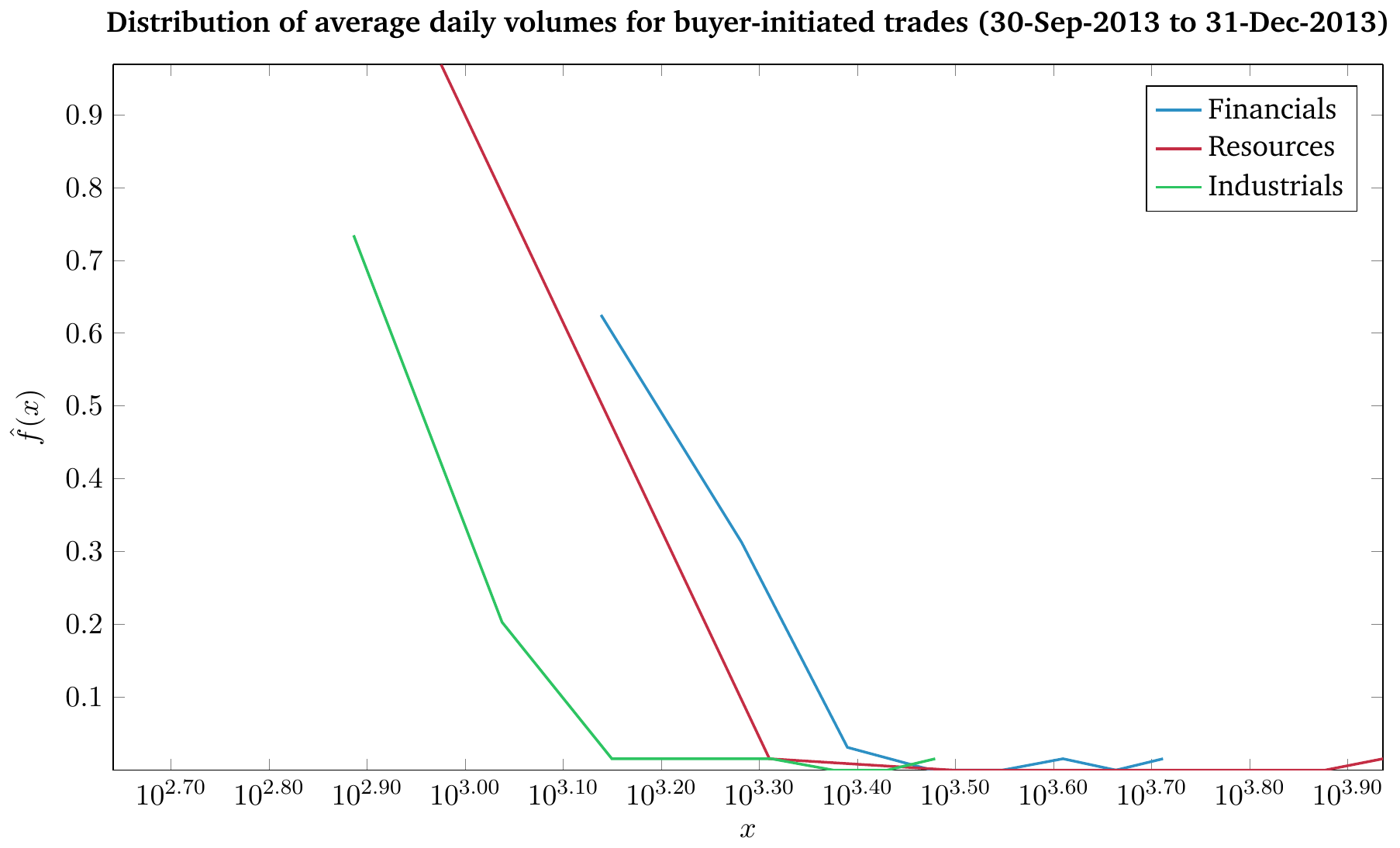}%
}
\caption{Plot of empirical distributions of daily average transaction volume for buyer initiated transactions of constituents of the Financials (JSE-FINI), Resources (JSE-RESI) and Industrials (JSE-INDI) sectors for the periods 01-Jan-2013 to  27-Sep-2013 (left) and 30-Sep-2013 to 31-Dec-2013 (right).}\label{EmpiDistBuyerVolumes}
\end{figure*}
\begin{figure*}[tbp!]
\centering
\subfloat[Period before fee structure change.]{%
\includegraphics[width=0.49\textwidth]{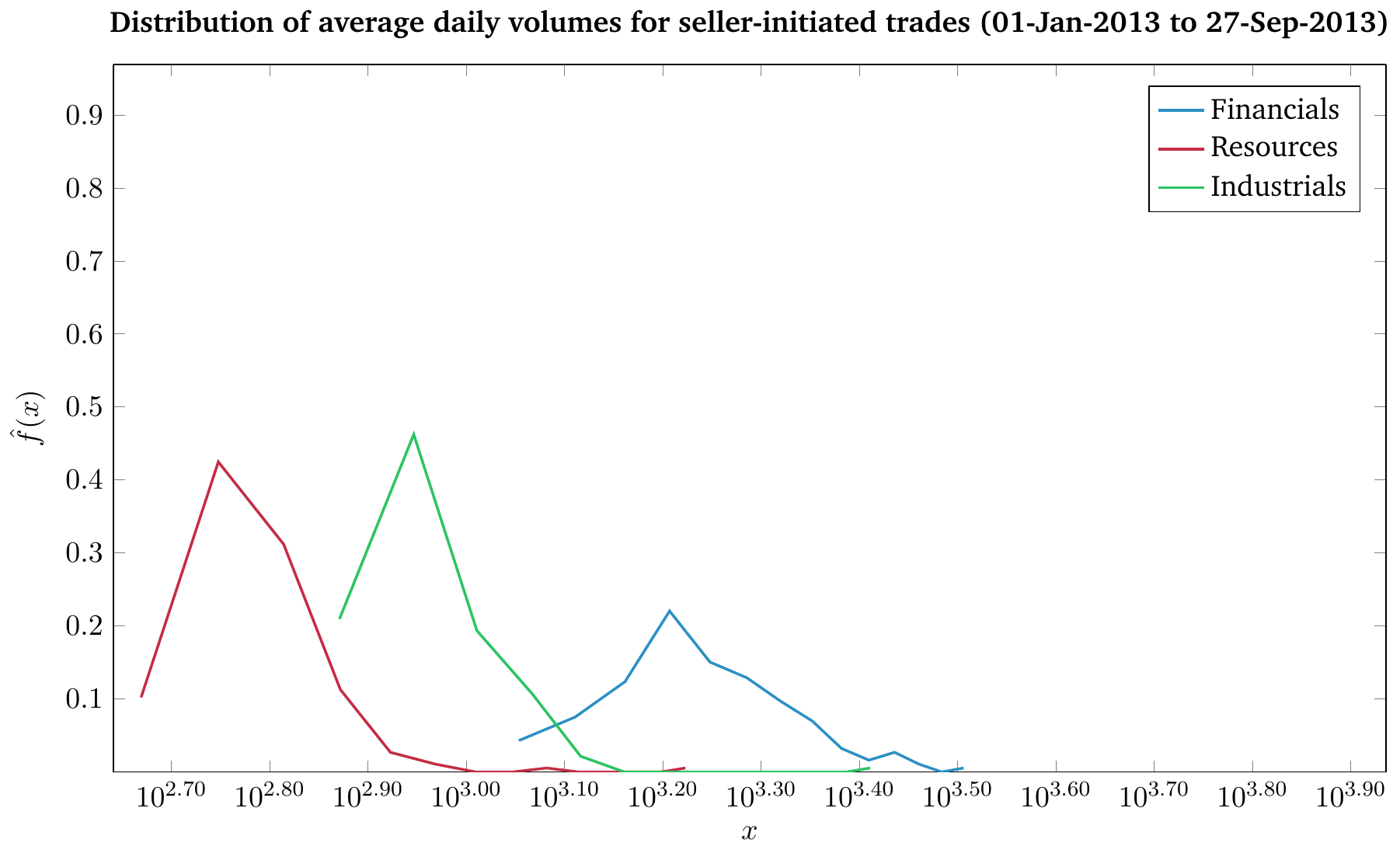}%
}
\hfill
\subfloat[Period after fee structure change.]{%
\includegraphics[width=0.49\textwidth]{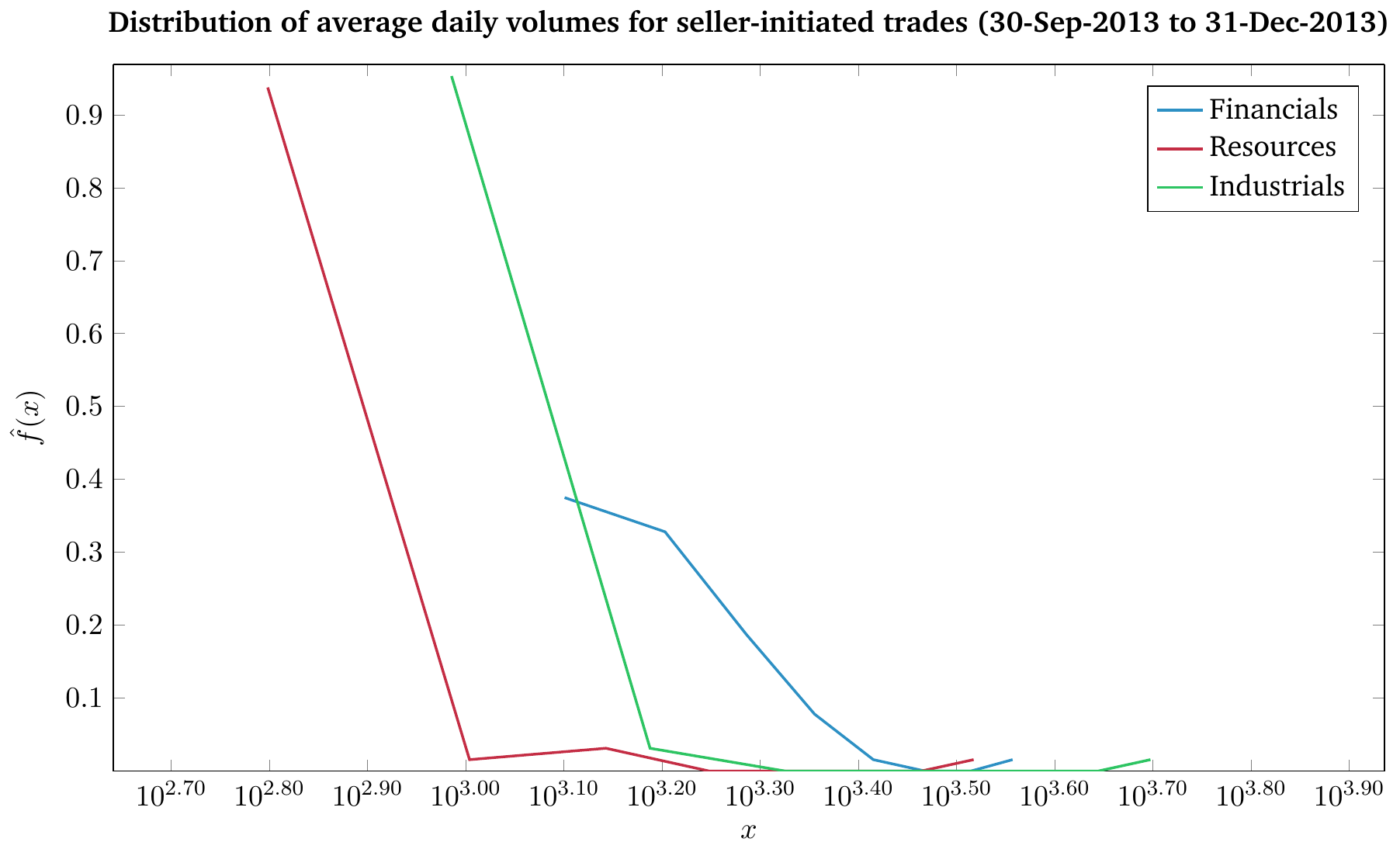}%
}
\caption{Plot of empirical distributions of daily average transaction volume for seller initiated transactions of constituents of the Financials (JSE-FINI), Resources (JSE-RESI) and Industrials (JSE-INDI) sectors for the periods 01-Jan-2013 to  27-Sep-2013 (left) and 30-Sep-2013 to 31-Dec-2013 (right).}\label{EmpiDistSellerVolumes}
\end{figure*}
\begin{figure*}[tbp!]
\centering
\subfloat[Period before fee structure change.]{%
\includegraphics[width=0.49\textwidth]{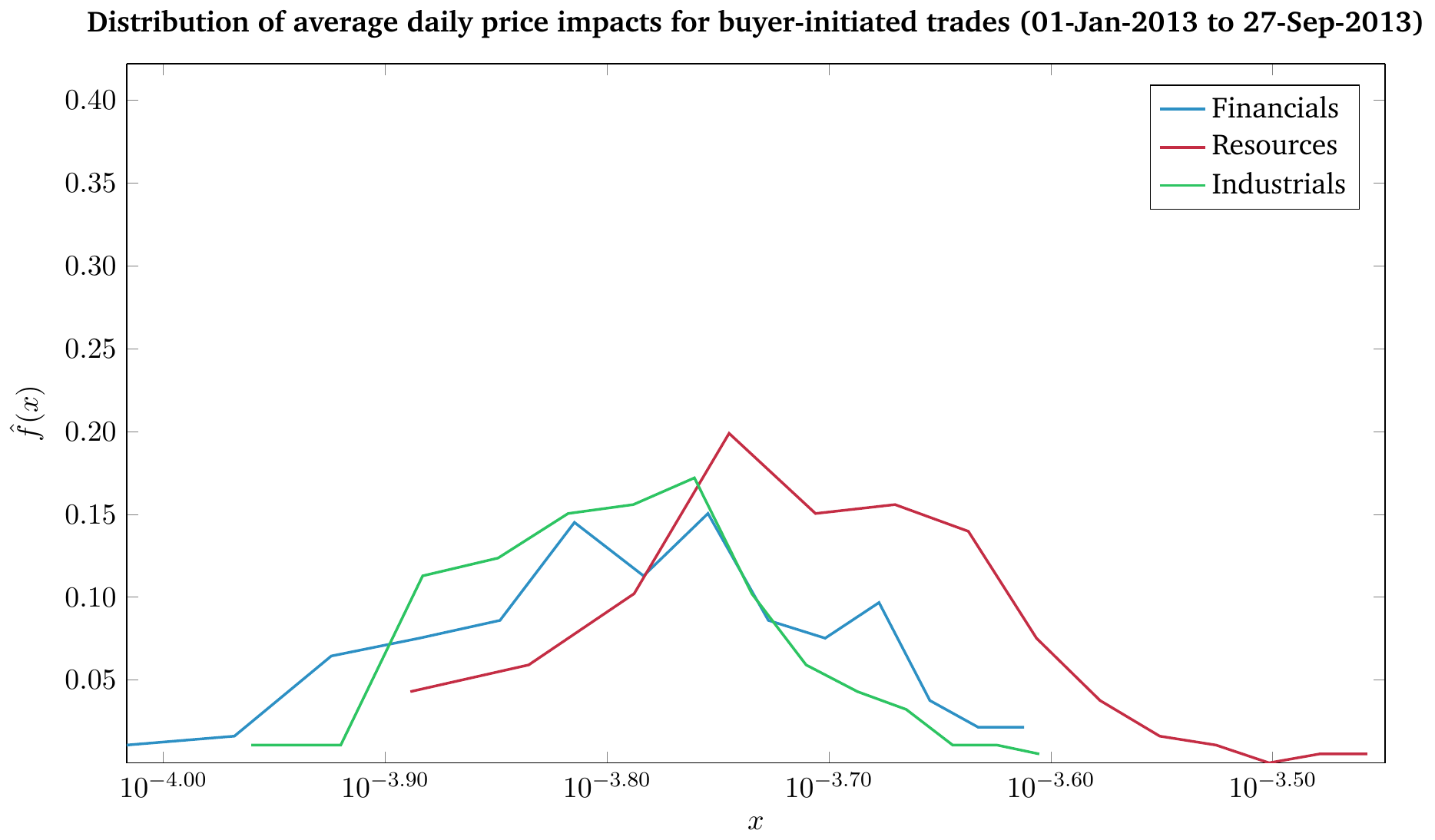}%
}
\hfill
\subfloat[Period after fee structure change.]{%
\includegraphics[width=0.49\textwidth]{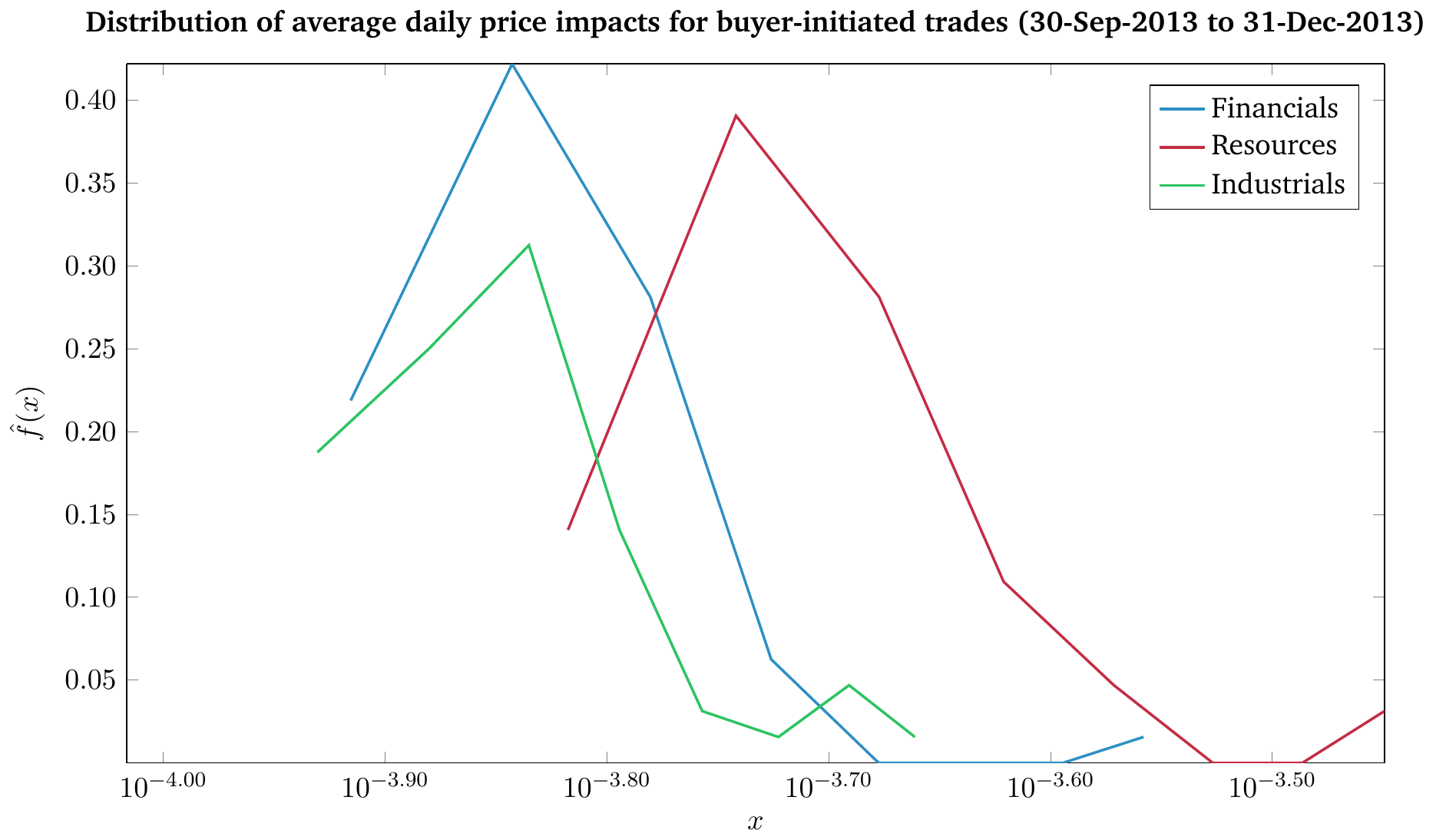}%
}
\caption{Plot of empirical distributions of daily average transaction price impact for buyer initiated transactions of constituents of the Financials (JSE-FINI), Resources (JSE-RESI) and Industrials (JSE-INDI) sectors for the periods 01-Jan-2013 to  27-Sep-2013 (left) and 30-Sep-2013 to 31-Dec-2013 (right).}\label{EmpiDistBuyerPriceImpacts}
\end{figure*}
\begin{figure*}[tbp!]
\centering
\subfloat[Period before fee structure change.]{%
\includegraphics[width=0.49\textwidth]{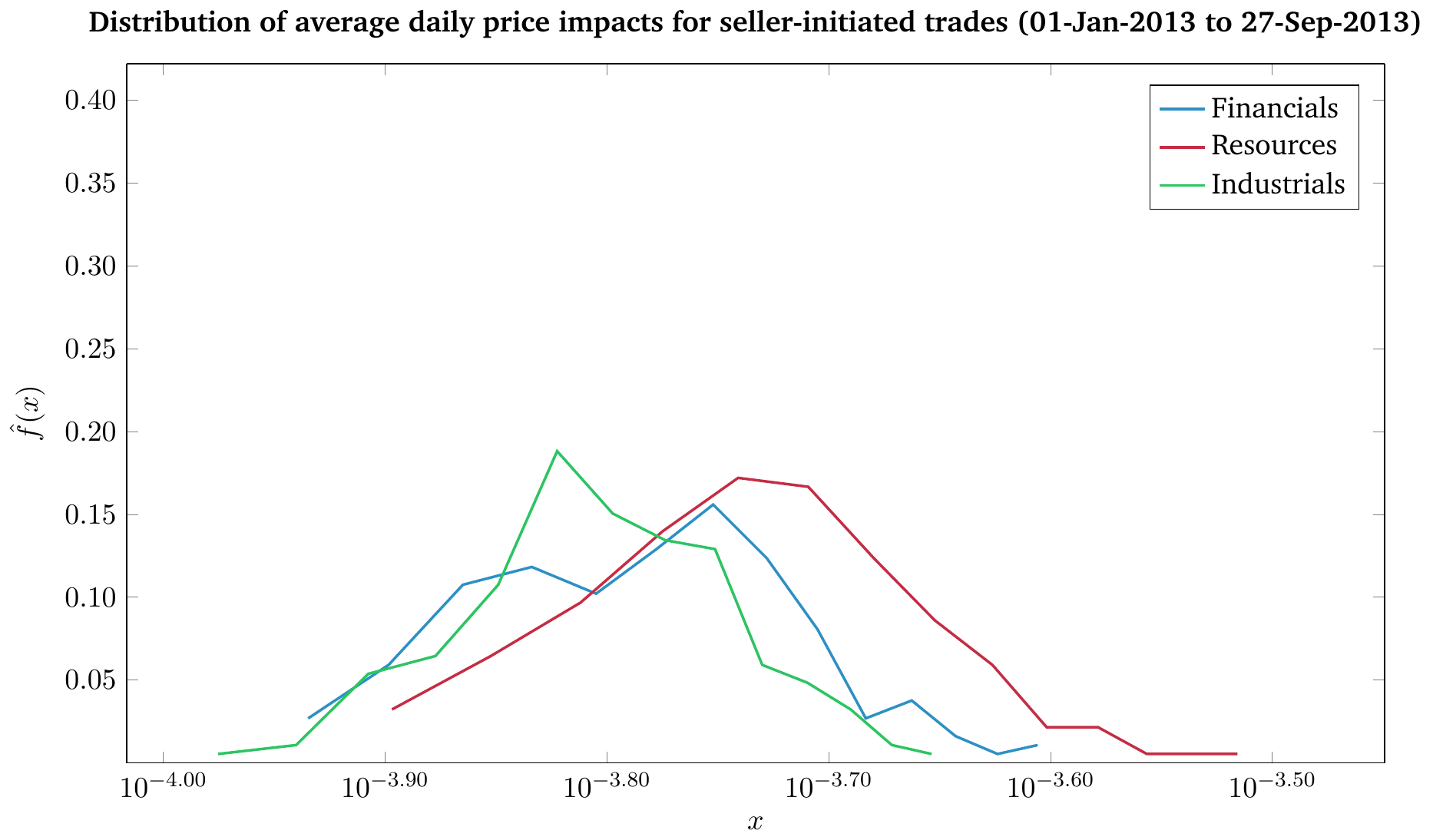}%
}
\hfill
\subfloat[Period after fee structure change.]{%
\includegraphics[width=0.49\textwidth]{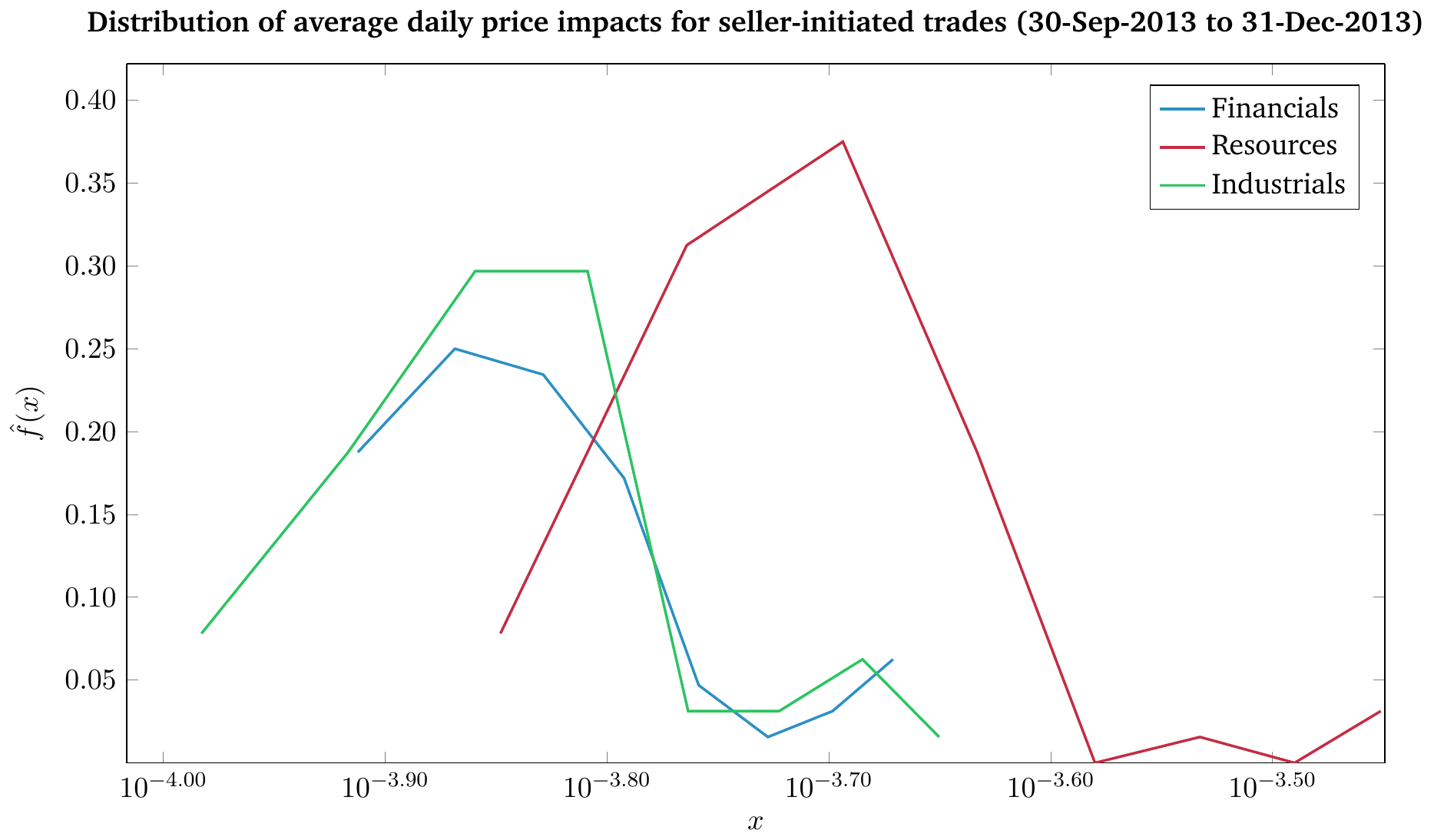}%
}
\caption{Plot of empirical distributions of daily average transaction price impact for seller initiated transactions of constituents of the Financials (JSE-FINI), Resources (JSE-RESI) and Industrials (JSE-INDI) sectors for the periods 01-Jan-2013 to  27-Sep-2013 (left) and 30-Sep-2013 to 31-Dec-2013 (right).}\label{EmpiDistSellerPriceImpacts}
\end{figure*}
\begin{figure*}[tbp!]
\centering
\subfloat[Period before fee structure change.]{%
\includegraphics[width=0.49\textwidth]{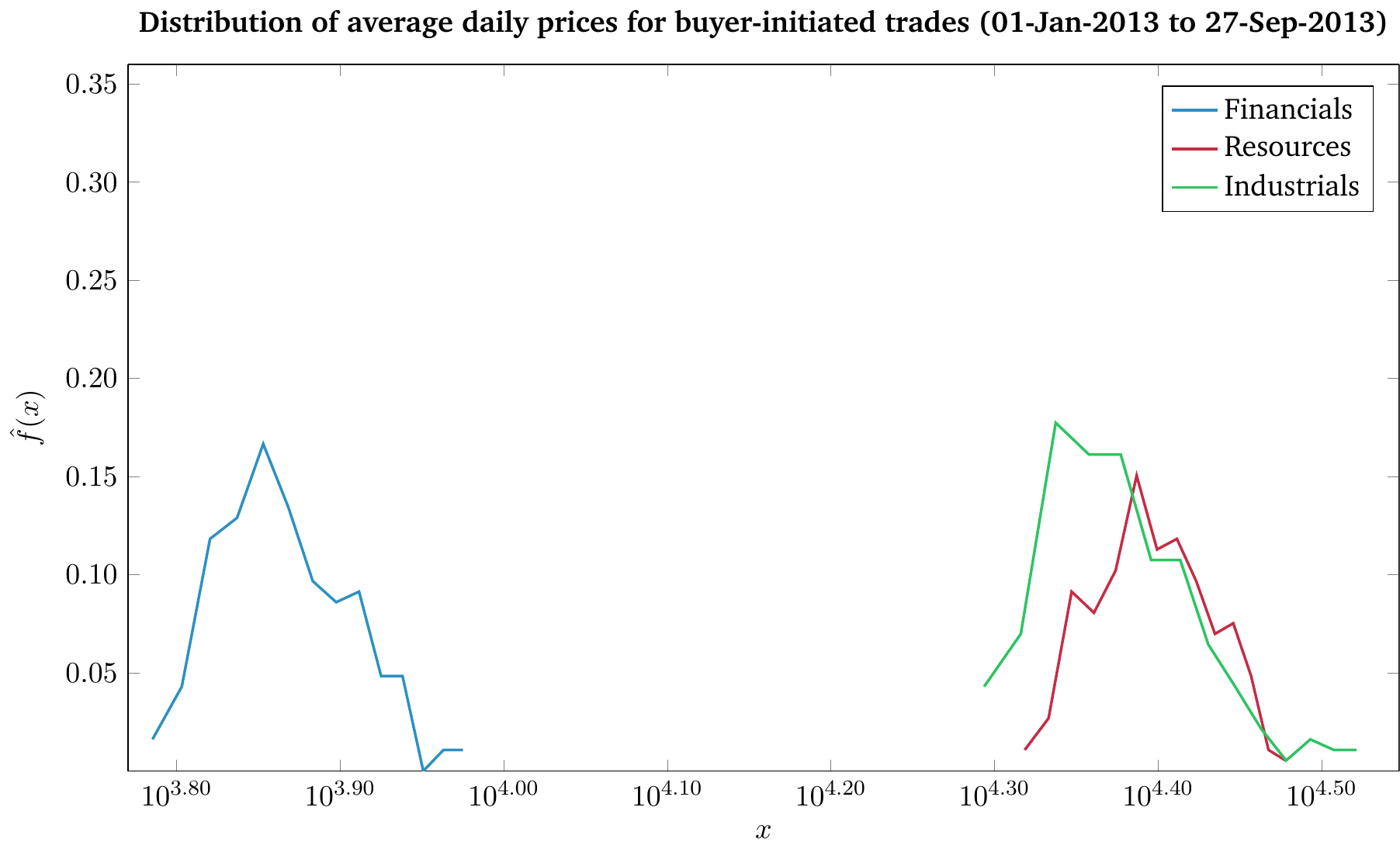}%
}
\hfill
\subfloat[Period after fee structure change.]{%
\includegraphics[width=0.49\textwidth]{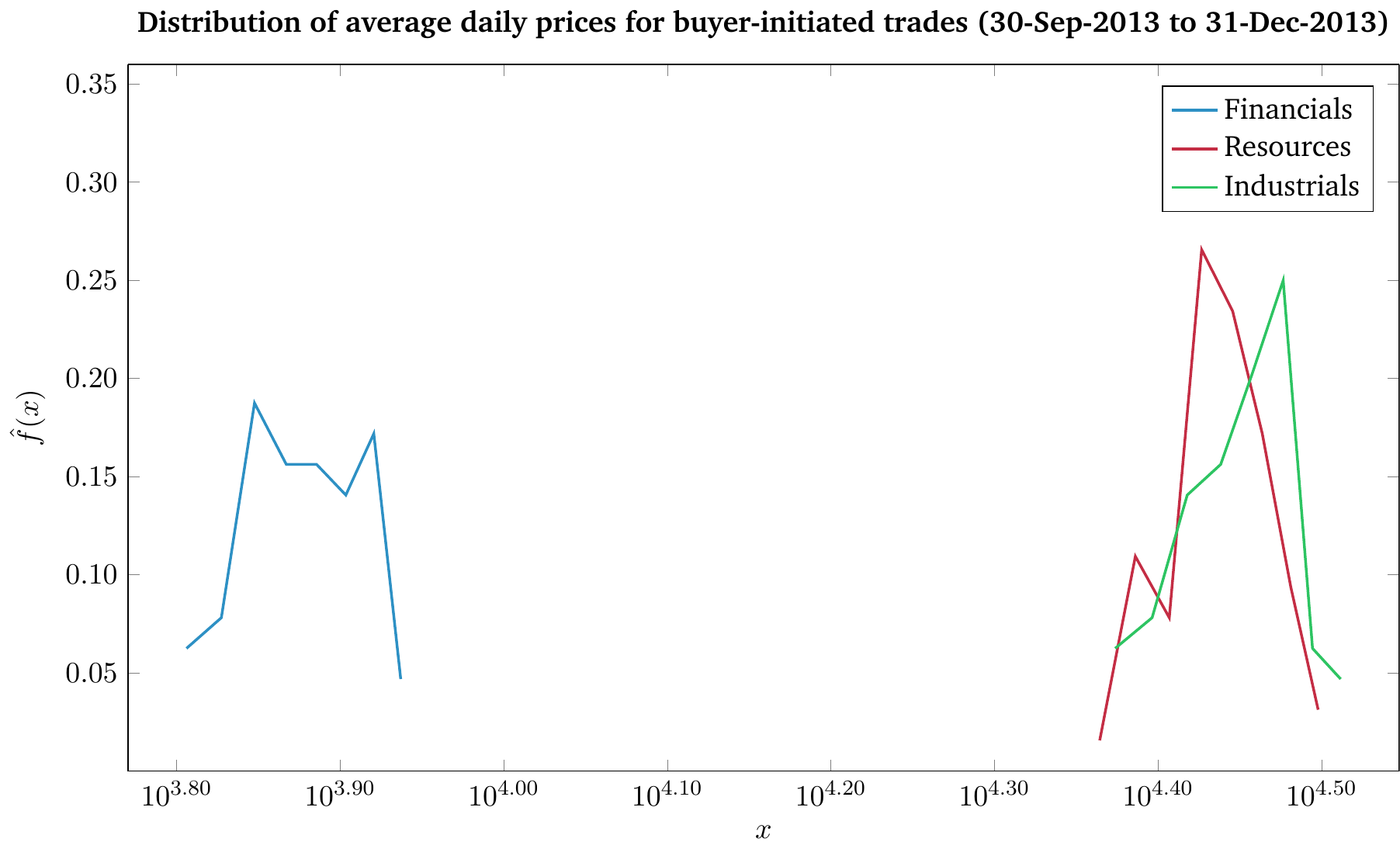}%
}
\caption{Plot of empirical distributions of daily average transaction price for buyer initiated transactions of constituents of the Financials (JSE-FINI), Resources (JSE-RESI) and Industrials (JSE-INDI) sectors for the periods 01-Jan-2013 to  27-Sep-2013 (left) and 30-Sep-2013 to 31-Dec-2013 (right).}\label{EmpiDistBuyerPrices}
\end{figure*}
\begin{figure*}[tbp!]
\centering
\subfloat[Period before fee structure change.]{%
\includegraphics[width=0.49\textwidth]{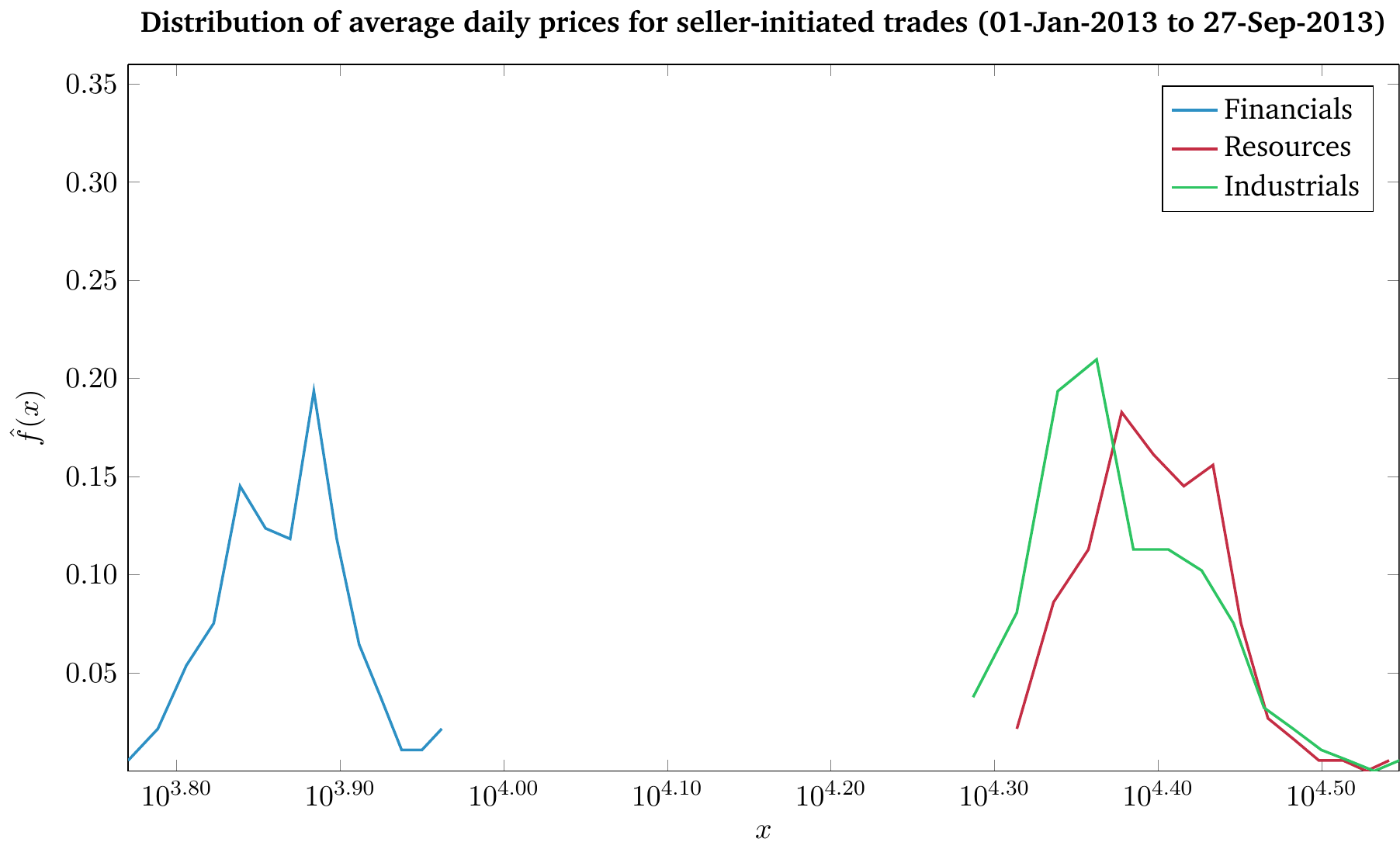}%
}
\hfill
\subfloat[Period after fee structure change.]{%
\includegraphics[width=0.49\textwidth]{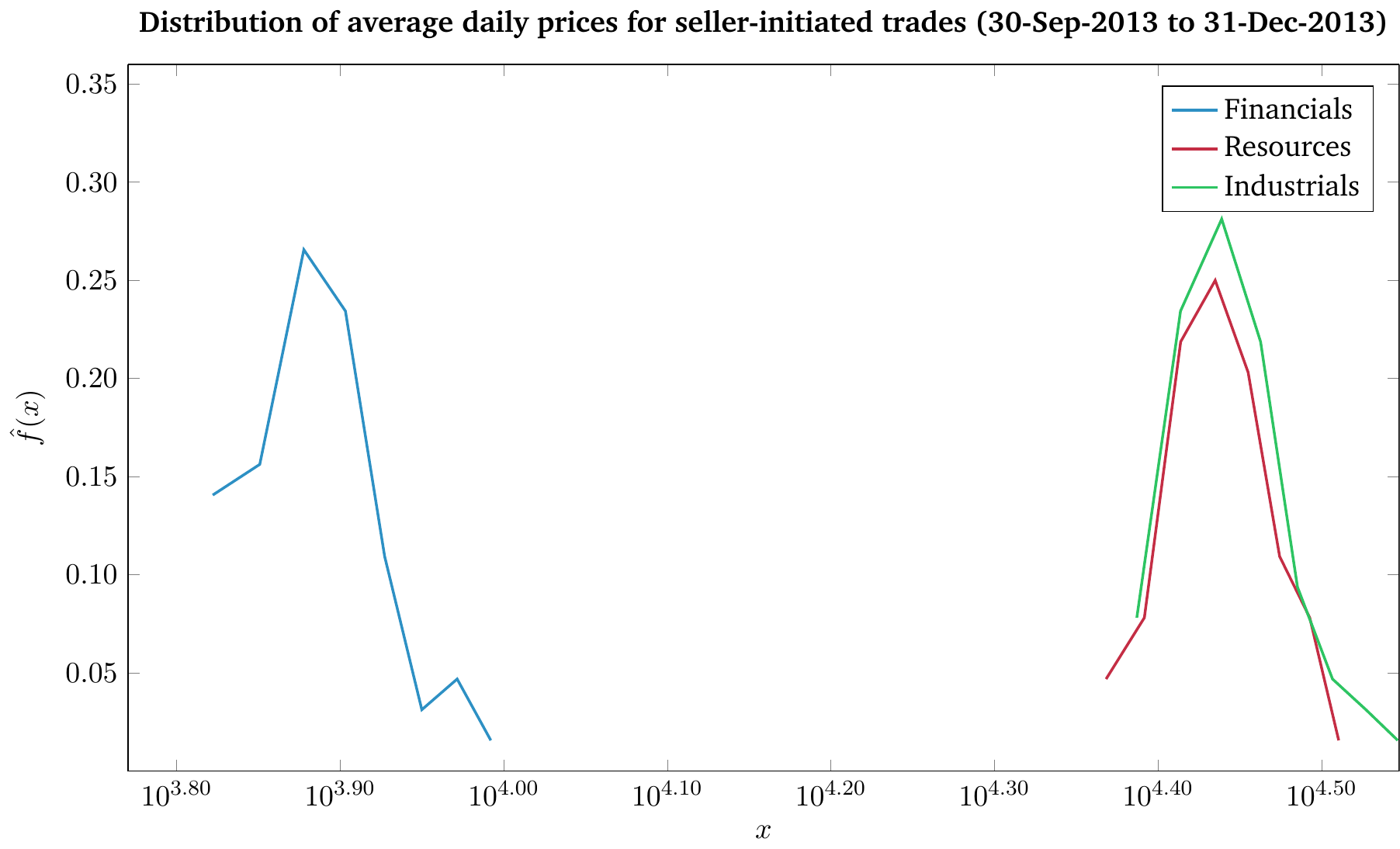}%
}
\caption{Plot of empirical distributions of daily average transaction price for seller initiated transactions of constituents of the Financials (JSE-FINI), Resources (JSE-RESI) and Industrials (JSE-INDI) sectors for the periods 01-Jan-2013 to  27-Sep-2013 (left) and 30-Sep-2013 to 31-Dec-2013 (right).}\label{EmpiDistSellerPrices}
\end{figure*}

\section{Power-law size distributions for price impact}\label{app:powerlaws}

\begin{table*}[tbp!]%
\caption{%
Summary of results of power-law size distribution fits to a subset of the average price impact data~$\left(\Delta p^{\ast}\right)$. Prior to fitting the power-law size distributions to the $20$ average price impact data points obtained from the logarithmic binning of transactions with normalised transaction volume bewteen $10^{-3.2}$ and $10$, a subset of the average price impact data are obtained by choosing a lower bound on the average normalised transaction volume  by inspecting the graphs in~\cref{MasterCurveBuyer,MasterCurveSeller}. Specifically, power-law size distributions are fit only to those average price impact data that have an average normalised transaction volume exceeding $10^{-0.9}$. This is done separately for both buyer and seller initiated trades for all three sectors studied for the periods before (01-Jan-2013 to 27-Sep-2013) and after (30-Sep-2013 to 31-Dec-2013) the fee structure change. The estimated scaling exponent $\left(\alpha\right)$, the lower bound of the power-law behaviour $\left(x_{min}\right)$ and the \textit{p}-value $\left(p_{value}\right)$ for goodness-of-fit are given separately for buyer and seller initiated transactions from each of the three sectors for periods before and after the fee structure change. The method (and corresponding MATLAB code) of Clauset, Shalizi and Newman (2009)~\cite{CSN2009}~was used.%
}\label{tab:PowerLawFitsTable}%
\centering%
\begin{tabular}{@{}L{0.142857\textwidth} L{0.142857\textwidth} L{0.142857\textwidth} C{0.142857\textwidth} C{0.142857\textwidth} C{0.142857\textwidth}@{}}%
\toprule%
\textbf{Sector} & %
\textbf{Direction} & %
\textbf{Period} & %
\textbf{$\alpha$} & %
\textbf{$x_{min}$} & %
\textbf{$p_{value}$} \\\midrule%
\multirow{4}{*}[-3.2pt]{Financials} & \multirow{2}{*}{Buyer}  & Before & $ 4.0249 $ & $ 1.4448 \times 10^{-4} $ & $ 0.1278 $ \\ %
                             &                         & After  & $ 4.0400 $ & $ 1.2475 \times 10^{-4} $ & $ 0.1033 $ \\ \cmidrule{2-6} %
                             & \multirow{2}{*}{Seller} & Before & $ 4.1053 $ & $ 1.4423 \times 10^{-4} $ & $ 0.1387 $ \\ %
                             &                         & After  & $ 4.4918 $ & $ 1.2856 \times 10^{-4} $ & $ 0.2012 $ \\ \midrule %
\multirow{4}{*}[-3.2pt]{Resources} & \multirow{2}{*}{Buyer}  & Before & $ 3.7082 $ & $ 1.6691 \times 10^{-4} $ & $ 0.3640 $ \\ %
                             &                         & After  & $ 3.1030 $ & $ 1.4396 \times 10^{-4} $ & $ 0.3620 $ \\ \cmidrule{2-6} %
                             & \multirow{2}{*}{Seller} & Before & $ 3.6888 $ & $ 1.5860 \times 10^{-4} $ & $ 0.4602 $ \\ %
                             &                         & After  & $ 3.7455 $ & $ 1.5638 \times 10^{-4} $ & $ 0.2645 $ \\ \midrule %
\multirow{4}{*}[-3.2pt]{Industrials} & \multirow{2}{*}{Buyer}  & Before & $ 3.6472 $ & $ 1.3919 \times 10^{-4} $ & $ 0.1043 $ \\ %
                             &                         & After  & $ 3.2715 $ & $ 1.1211 \times 10^{-4} $ & $ 0.2371 $ \\ \cmidrule{2-6} %
                             & \multirow{2}{*}{Seller} & Before & $ 3.4305 $ & $ 1.3791 \times 10^{-4} $ & $ 0.1239 $ \\ %
                             &                         & After  & $ 3.8937 $ & $ 1.2007 \times 10^{-4} $ & $ 0.2387 $ \\ \bottomrule%
\end{tabular}%
\end{table*}%

Power-law size distributions are concerned with how the frequency of an event relates to some measure of the size of that event. In mathematical terms, the quantity $x$ will obey a power-law if its probability is proportional to a power of itself~\cite{CSN2009}:
\[
p(x) \propto x^{-\alpha}.
\]
The constant $\alpha$ is commonly referred to as the \emph{scaling exponent} and it quantifies the strength of the linear relationship on a log-log scale.~%
We are specifically interested in considering if the frequency of price impact events vary as a power of price impact itself.

For each of the three sectors studied the method (and corresponding MATLAB code) of Clauset, Shalizi and Newman (2009) (as discussed in~\cite{CSN2009}) was used to fit power-law size distributions to a subset of the average price impact data of buyer and seller initiated transactions. The results for the fits of the power-law size distributions are presented in~\cref{tab:PowerLawFitsTable} where the estimated scaling exponent $\left(\alpha\right)$, the lower bound of the power-law behaviour $\left(x_{min}\right)$ and the \textit{p}-value $\left(p_{value}\right)$ for goodness-of-fit are given separately for buyer and seller initiated transactions from each of the three sectors for periods before and after the fee structure change.~%
For all sectors, trade directions and periods the estimated \textit{p}-values indicate a failure to reject the null hypothesis that price impact follows a power-law distribution at the $10\%$ significance level. Thus there is not enough statistical evidence to conclude that a power-law size distribution for price impact is not valid.

Slight changes in the scaling exponents of the power-law fits were observed across all sectors and trade directions when comparing the power-law behaviour for the periods before and after the fee structure change.~%
Increases in scaling exponents correspond to shifts in probability mass away from tail events and increases in the lower bounds of fits indicate that the power-law distribution begins further into the tail.

The fact that the estimated scaling exponents as well as the lower bounds on the power-law behaviour of statistically significant power-law fits vary before and after the fee structure change is indicative of a structural change in price impact, however one cannot say that the fee structure change resulted in this structural change.

\section{Sector constituents}\label{app:sectorconstituents}

The constituents of the three major sectors studied are given below. The Reuters Instrument Code, or RIC, is listed for each stock.~%

\vspace{0.3cm}\scriptsize

\noindent{\bf Financials / JSE-FINI (J212)}\\[0.1cm]
\noindent%
Discovery Holdings Ltd (DSYJ.J); %
Firstrand Ltd (FSRJ.J); %
Investec Ltd (INLJ.J); %
Investec PLC (INPJ.J); %
Nedbank Group Ltd (NEDJ.J); %
Old Mutual PLC (OMLJ.J); %
RMB Holdings Ltd (RMHJ.J); %
Sanlam Ltd (SLMJ.J); %
Standard Bank Group Ltd (SBKJ.J). %

\vspace{0.3cm}

\noindent{\bf Resources / JSE-RESI (J210)}\\[0.1cm]
\noindent%
African Rainbow Minerals Ltd (ARIJ.J); %
Anglo American Platinum Ltd (AMSJ.J); %
Anglo American PLC (AGLJ.J); %
AngloGold Ashanti Ltd (ANGJ.J); %
Assore Ltd (ASRJ.J); %
BHP Billiton PLC (BILJ.J); %
Gold Fields Ltd (GFIJ.J); 
Impala Platinum Holdings Ltd (IMPJ.J); %
Kumba Iron Ore Ltd (KIOJ.J); %
Mondi Ltd (MNDJ.J); %
Mondi PLC (MNPJ.J); %
Sasol Ltd (SOLJ.J).

\vspace{0.3cm}

\noindent{\bf Industrials / JSE-INDI (J211)}\\[0.1cm]
\noindent%
Aspen Pharmacare Holdings Ltd (APNJ.J); %
Bidvest Group Ltd (BVTJ.J); %
British American Tobacco PLC (BTIJ.J); %
Compagnie Financiere Richemont SA (CFRJ.J); %
Exxaro Resources Ltd (EXXJ.J); %
Growthpoint Properties Ltd (GRTJ.J); %
Imperial Holdings Ltd (IPLJ.J); %
Massmart Holdings Ltd (MSMJ.J); %
Mediclinic International Ltd (MDCJ.J); %
MTN Group Ltd (MTNJ.J); %
Naspers Ltd (NPNJn.J); %
Remgro Ltd (REMJ.J); %
SABMiller PLC (SABJ.J); %
Shoprite Holdings Ltd (SHPJ.J); %
Steinhoff International Holdings (SHFJ.J); %
Tiger Brands Ltd (TBSJ.J); %
Truworths International Ltd (TRUJ.J); %
Vodacom Group Ltd (VODJ.J); %
Woolworths Holdings Ltd (WHLJ.J).

\end{document}